\documentclass[epsfig,12pt,onecolumn]{article}
\usepackage{makeidx}
\usepackage{amsfonts}
\usepackage{amssymb}
\usepackage{amsmath}
\usepackage{multicol}
\usepackage{graphicx}
\usepackage{float}
\usepackage{caption}

\makeatletter \@addtoreset{equation}{section}
\renewcommand{\theequation}{\thesection.\arabic{equation}}
\def \be{\begin{equation}}
\def \ee{\end{equation}}
\def \bea{\begin{eqnarray}}
\def \eea{\end{eqnarray}}

\newcommand{\nc}{\newcommand}
\nc{\al}{\alpha} \nc{\bib}{\bibitem} \nc{\la}{\lambda}
\nc{\C}{\mbox{\hspace{1.24mm}\rule{0.2mm}{2.5mm}\hspace{-2.7mm} C}}
\nc{\R}{\mbox{\hspace{.04mm}\rule{0.2mm}{2.8mm}\hspace{-1.5mm} R}}
\input{tcilatex}
\begin{document}

\title{\textbf{Minuscule ABCDE Lax Operators from }\\
\textbf{4D Chern-Simons Theory}}
\author{Youssra Boujakhrout and El Hassan Saidi \\
{\small 1. LPHE-MS, Science Faculty}, {\small Mohammed V University in
Rabat, Morocco}\\
{\small 2. Centre of Physics and Mathematics, CPM- Morocco}}
\maketitle

\begin{abstract}
Using 4D Chern-Simons (CS) theory with gauge symmetry $G$ having minuscule
coweights, we develop a suitable operator basis to deal with the explicit
calculation of the Lax operator of integrable spin chain satisfying the RLL
equation. Using this basis, we derive the oscillator realisations of the
full list of the minuscule L-operators which are classified by the gauge
symmetries A$_{N}$, B$_{N}$, C$_{N}$, D$_{N}$, E$_{6}$, E$_{7}$. We also
complete missing results regarding the non simply laced $SO_{2N+1}$ and $%
SP_{2N}$ gauge symmetries and comment on their intrinsic features. Moreover,
we investigate the properties of links reported in Yangian spin chain
studies between the (A-,D-) Lax operators and the (C-,B-) homologue. We show
that these links are due to discrete outer-automorphism symmetries that are
explicitly worked out. \  \newline
\  \  \  \  \  \  \newline
\textbf{Keywords:} 4D CS theory, Crossing Wilson and 't Hooft lines,
Yang-Baxter and RLL equations. Quantum integrability, Oscillator realisation
of Lax operator.
\end{abstract}


\section{ Introduction}

Few years ago, a four dimensional (4D) topological gauge theory with
complexified gauge symmetry G \textrm{\cite{1A,2A}} has been proposed to be
a mother theory of lower dimensional integrable systems such as quantum 1D
integrable spin chains of statistical mechanics \textrm{\cite{1B}-\cite{4B}}
and 2D integrable QFTs \textrm{\cite{10B}-\cite{15B}}. This 4D topological
gauge theory is a tricky extension of the usual non abelian 3D Chern-Simons
theory \textrm{\cite{1C}}\ with observables given by topological line
defects such as the electrically charged Wilson lines and the magnetically
charged 't Hooft lines. The 4D Chern-Simons (CS) theory has been linked with
$\mathcal{N}=(1,1)$ supersymmetric Yang Mills theory in 6D \textrm{\cite{20A}%
-\cite{20C}} and supersymmetric quiver gauge theories \cite{20D}-\cite{20H}.
The observables of the 4D CS theory have been also realised in terms of
M2/M5-branes and also in terms of intersecting NS5/Dp- branes in type II
strings \textrm{\cite{21}-\cite{28}}. \newline
Recently, it has been shown in \textrm{\cite{4DA,4D}} that the coupling\ of
a 't Hooft line with N perpendicular Wilson lines in the four-dimensional
theory generates an integrable quantum spin chain \textrm{\cite{1D}}. The
crossing matrix describing the coupling of an electrically charged Wilson
and a magnetically charged 't Hooft line coincides precisely with the Lax-
operator $\mathcal{L}$ of integrable systems \textrm{\cite{2D}}. In this 4D
Chern-Simons (CS) setup, a beautiful formula for calculating the crossing
matrix $\mathcal{L}$ has been derived for the sub-family of 't Hooft lines
whose magnetic charges are given by minuscule coweights $\mu $ of the gauge
symmetry G. In these regards, well known results from the literature of
quantum spin chains \textrm{\cite{3DA,3DB}} have been nicely recovered from
the point of view of the 4D Chern-Simons theory and partial findings
including QFT modelings and their interpretation have been completed \textrm{%
\cite{4D}-\cite{6D}}. In fact, the oscillator realisation of Lax matrices of
A- and D-types derived from CS theory solving the RLL equations were shown
to reproduce, up to a spectral parameter scaling, their homologue derived by
using analytic Yangian based methods \textrm{\cite{0E,1E}}. Moreover, new
results such as the $\mathcal{L}$-operators for minuscule representations of
the\ exceptional symmetries $E_{6}$\ and $E_{7}$ were generated from the CS
gauge theory analysis in\  \textrm{\cite{1G}}. Notice that the derivation of
Lax-matrices for the exceptional groups from the algebraic integrability
methods is still missing in the quantum integrable spin chain literature
using Yangian algebra. \newline
\textrm{In this paper}, we contribute to the 4D Chern-Simons theory with two
main objectives. The first objective aims at $\left( \mathbf{i}\right) $ the
development of a suitable method for the explicit computation of the
L-operator of integrable systems; and $\left( \mathbf{ii}\right) $ to
complete partial results in the study of Lax operators in the framework of
the CS theory with gauge symmetry G. A particular interest is also devoted
to the missing analysis concerning the CS gauge theory with non simply laced
B- and C- gauge symmetries and to the derivation of the associated harmonic
oscillator representation of the Lax operators $\mathcal{L}_{B}$ and $%
\mathcal{L}_{C}$. Moreover, by using the 4D CS approach, we identify the
source behind similarities reported in the integrable spin chain literature
between B- and D- Lax operators and between C- and A- types Lax operators.
These links are shown to be due to discrete symmetries of the concerned $%
\mathcal{L}_{G}$'s whose cause is given by known outer-automorphism
symmetries of the root system of the gauge symmetry G. We also show that our
results based on CS theory are in perfect agreement with the quantum spin
chains' calculations obtained recently using Yangian algebra \textrm{\cite%
{1E}}.\newline
The second objective is to calculate the complete list of all minuscule Lax
operators $\mathcal{L}_{G}.$ This list is presented into a unified set (see
Tables \textbf{\ref{T2},} \textbf{\ref{T4}} and section 3) in order to
provide to the interested reader a summary tool on the $\mathcal{L}_{G}$'s
for easy use and also for a suitable parametrisation for further
applications. In this regard, we recall that the minuscule Lax operators as
formulated in the 4D Chern-Simons theory are classified by gauge symmetries
G given by the series A$_{N}$, B$_{N}$, C$_{N}$, D$_{N}$, and the
exceptional E$_{6}$ and E$_{7}$. For a unified description of the operators $%
\mathcal{L}_{G}$, we revisit the explicit construction of the harmonic
oscillator realisations of these $\mathcal{L}_{G}$'s and we investigate
their properties and the relationships outlined in literature. \newline
\textrm{The organisation of this paper} is as follows: In section 2, we
revisit aspects regarding the Lax-operator of integrable systems from two
points of view: First, from the view of integrable spin chains method, as a
matter of completeness; and second from the novel view of 4D CS theory.
\textrm{We also indicate the main way to follow in order to reach our goals}%
. In section 3, we give the full list of minuscule $\mathcal{L}_{G}$-
operators calculated from the 4D Chern-Simons theory with gauge symmetry G.
This list concerns the minuscule A$_{N}$, B$_{N}$, C$_{N}$, D$_{N}$ families
and the exceptional E$_{6/7}$ with electrically charged Wilson lines taken
in the fundamental vector-like representation. Line in spinor-like
representation are investigated in the discussion section. For unified
notations, we revisit all the calculations by using standard writings of
Hilbert quantum states. In section 4, we give the explicit derivation of the
$\mathcal{L}_{B}$- operator of the non simply laced SO$_{2N+1}$gauge
symmetry and give a comparison with the $\mathcal{L}_{D}$- operator of the
simply laced SO$_{2N}$ theory. In section 5, we derive the $\mathcal{L}_{C}$%
- operator of the non simply laced SP$_{2N}$ series and comment on its links
with the $\mathcal{L}$- operators of the simply laced SO$_{2N}$- and SL$%
_{2N} $- gauge theories. Section 6 is devoted to conclusion and discussions.

\section{Lax operator families in integrable systems}

\begin{figure}[tbph]
\begin{center}
\includegraphics[width=4cm]{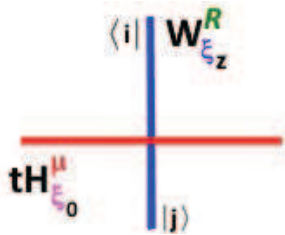}
\end{center}
\par
\vspace{-0.5cm}
\caption{The operator $\mathcal{L}_{G}$ describing the matrix coupling
between a horizontal 't Hooft line (tH$_{\protect \xi _{0}}^{\protect \mu }$
in red) at z=0 and a vertical Wilson line (W$_{\protect \xi _{z}}^{R}$ in
blue) at z with incoming $\left \langle i\right \vert $ and out going $%
\left \vert j\right \rangle $ states. The spectral parameter z is interpreted
in 4D CS theory as a point in the holomorphic plane.}
\label{WH}
\end{figure}

In this section, we revisit the construction of minuscule Lax operators $%
\mathcal{L}_{G}$ for low dimensional integrable systems with symmetry Lie
group G and refine aspects towards $\mathcal{L}_{G}$. We recall basic tools
on ABCDE symmetries which are used in the forthcoming sections and also
investigate the way to the minuscule $\mathcal{L}_{G}$'s. These L-operators
are classified by the symmetry groups G having minuscule coweights. Their
matrix representations have an interpretation as a matrix couplings of two
topological lines as shown by the Figure \textbf{\ref{WH}. }

As one of the main objective of this study is to give the full list of $%
\mathcal{L}_{G}$s, we think it interesting to: $\left( \mathbf{1}\right) $
describe here those basic quantities by using physical language and short
paths for their properties. $\left( \mathbf{2}\right) $ emphasize the power
of the QFT method of \textrm{\cite{1A,2A} compared to the standard}
algebraic approach based on the Yangian algebra representations \cite{1F}. $%
\left( \mathbf{3}\right) $ draw an overview on the minuscule operators $%
\mathcal{L}_{G}$ and anticipate the construction by giving some results that
will be derived rigorously later in this paper.

\subsection{Families of minuscule $\mathcal{L}$-operators}

\begin{figure}[tbph]
\begin{center}
\includegraphics[width=9cm]{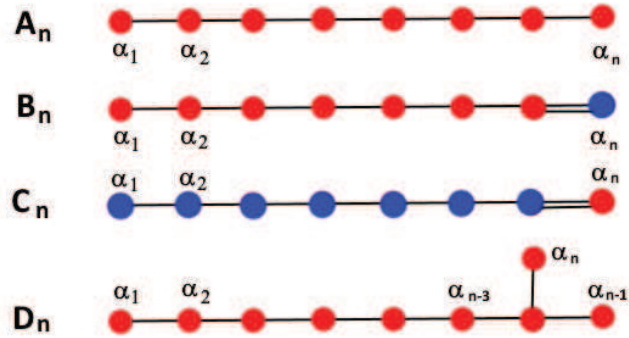}
\end{center}
\par
\vspace{-0.5cm}
\caption{Classification of minuscule Lax operators in terms of the Dynkin
diagrams of simple groups with minuscule coweights. Here, we have given the
families $A_{n}=sl_{n+1},$ $B_{n}=so_{2n+1},$ $C_{n}=sp_{2n}$ and $%
D_{n}=so_{2n}$. }
\label{DD}
\end{figure}

\begin{figure}[tbph]
\begin{center}
\includegraphics[width=10cm]{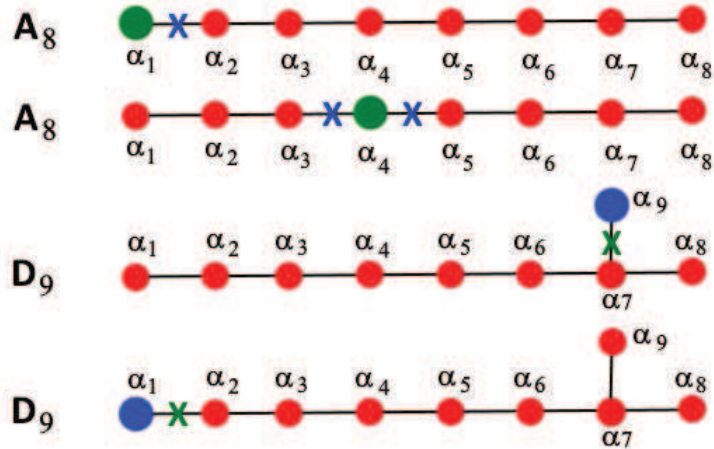}
\end{center}
\par
\vspace{-0.5cm}
\caption{Four examples of graphs describing minuscule Lax operators. The
first example concerns cutting the first node in A$_{8}$. The second example
regards the cutting of the fourth minuscule node in A$_{8}$. These two
graphs describe two different Lax operators. The third and the fourth graphs
deals with Lax operators classified by D$_{9}$.}
\label{LG}
\end{figure}

We begin by recalling that, in addition to the symmetry Lie group G, the
minuscule Lax-operators $\mathcal{L}_{G}$ of integrable 1D quantum spin
chains and 2D integrable QFT systems are also classified by the minuscule
coweights $\mu $ of finite dimensional Lie algebras. These minuscule
coweights are quite well known in the mathematical literature on Lie group
symmetries \textrm{\cite{2F}; }their useful properties for physical
applications are as collected in the following Table \textbf{\ref{T1}.}
\begin{table}[h]
\centering \renewcommand{\arraystretch}{1.2} $%
\begin{tabular}{|c|c|c|c|c|}
\hline
$G$ & minuscule$\  \mu $ & Levi algebra $l_{\mu }$ & $\dim $Rep & nilpotent$\
\ n_{\pm }$ \\ \hline \hline
$A_{n}$ & $\left.
\begin{array}{c}
\mu _{1} \\
\vdots \\
\mu _{k} \\
\vdots \\
\mu _{n}%
\end{array}%
\right. $ & $\left.
\begin{array}{c}
sl_{1}\oplus A_{n-1} \\
\vdots \\
sl_{1}\oplus A_{k}\oplus A_{n-k} \\
\vdots \\
sl_{1}\oplus A_{n-1}%
\end{array}%
\right. $ & $\left.
\begin{array}{c}
n+1 \\
\vdots \\
\frac{\left( n+1\right) !}{k!\left( n+1-k\right) !} \\
\vdots \\
n+1%
\end{array}%
\right. $ & $\left.
\begin{array}{c}
n_{\pm } \\
\vdots \\
k(n+1-k)_{\pm } \\
\vdots \\
n_{\pm }%
\end{array}%
\right. $ \\ \hline
$B_{n}$ & $\mu _{1}$ & so$_{2}\oplus B_{n-1}$ & $2n$ & $(2n-1)_{\pm }$ \\
\hline
$C_{n}$ & $\mu _{n}$ & so$_{2}\oplus A_{n-1}$ & $2^{n}$ & $\frac{1}{2}%
n(n+1)_{\pm }$ \\ \hline
$D_{n}$ & $\left.
\begin{array}{c}
\mu _{1} \\
\mu _{n-1} \\
\mu _{n}%
\end{array}%
\right. $ & $\left.
\begin{array}{c}
so_{2}\oplus D_{n-1} \\
so_{2}\oplus A_{n-1} \\
so_{2}\oplus A_{n-1}^{\prime }%
\end{array}%
\right. $ & $\left.
\begin{array}{c}
2n \\
2^{n-1} \\
2^{n-1}%
\end{array}%
\right. $ & $\left.
\begin{array}{c}
(2n-2)_{\pm } \\
\frac{1}{2}n(n-1)_{\pm } \\
\frac{1}{2}n(n-1)_{\pm }%
\end{array}%
\right. $ \\ \hline
$E_{6}$ & $\left.
\begin{array}{c}
\mu _{1} \\
\mu _{6}%
\end{array}%
\right. $ & $so_{2}\oplus D_{5}$ & $\left.
\begin{array}{c}
27 \\
\overline{27}%
\end{array}%
\right. $ & $\left.
\begin{array}{c}
16_{\pm } \\
16_{\pm }%
\end{array}%
\right. $ \\ \hline
$E_{7}$ & $\mu _{1}$ & $so_{2}\oplus E_{6}$ & $56$ & $27_{\pm }$ \\ \hline
\end{tabular}%
$%
\caption{The list of the fundamental minuscule coweights $\protect \mu $ of
finite dimensional Lie algebras g. The Levi-decomposition of these algebras
g with respect to the minuscule coweights is given by $\boldsymbol{n}%
_{-}\oplus \boldsymbol{l}_{\protect \mu }\oplus \boldsymbol{n}_{+}$. Here,
the $\boldsymbol{l}_{\protect \mu }$ refers to the Levi-subalgebra and $%
\boldsymbol{n}_{\pm }$\ to the nilpotent subalgebras.\ }
\label{T1}
\end{table}
In this classification table, we have also reported the Levi-subalgebra $%
\boldsymbol{l}_{\mu }$ associated with the Lie algebra g underlying the
symmetry group G. As well, we have given the nilpotent $\boldsymbol{n}_{\pm
} $ subalgebras accompanying $\boldsymbol{l}_{\mu }$ and which turn out play
an important role in the construction of the $\mathcal{L}_{G}^{\mu }$'s.
From this global table, we learn amongst others that there exist five
families of minuscule operators $\mathcal{L}_{G}^{\mu }$ denoted below as
follows%
\begin{equation}
\begin{tabular}{lllll}
$\mathcal{L}_{A}^{\mu },$ & $\mathcal{L}_{B}^{\mu },$ & $\mathcal{L}%
_{C}^{\mu },$ & $\mathcal{L}_{D}^{\mu },$ & $\mathcal{L}_{E}^{\mu }$%
\end{tabular}%
\end{equation}%
As exhibited on the Table \textbf{\ref{T1}}, the four first operators $%
\mathcal{L}_{G}^{\mu }$ constitute four infinite families $\left \{ \mathcal{%
L}_{G_{n}}^{\mu }\right \} $ labeled by the positive integer n (rank of G)
and by the minuscule coweight $\mu $ of G. The fifth family in the Table
\textbf{\ref{T1}} is finite, it concerns the exceptional $\mathcal{L}%
_{E_{6}}^{\mu }$ and $\mathcal{L}_{E_{7}}^{\mu }$. Because a given family
may have more than one minuscule coweight; say $\mu _{k}$ labeled by an
integer k, it results that the classification of the minuscule L-operators
is given by two integers: $\left( \mathbf{i}\right) $ The rank $n_{G}$ of
the gauge symmetry G$_{n}$ with Lie algebra g$_{n}$. $\left( \mathbf{ii}%
\right) $ The number n$_{\mu }$ of minuscule coweights $\mu _{k}$ for each
gauge symmetry G$_{n}$.\  \newline
Moreover, knowing that the minuscule coweights $\mu _{i}$ of finite Lie
algebras are intimately related with their simple roots $\alpha _{i}$ as
shown by the following duality relation
\begin{equation}
\mu _{i}.\alpha _{j}=\delta _{ij}
\end{equation}%
it follows that the $\mathcal{L}_{G}^{\mu }$'s can be put in correspondence
with the Dynkin diagrams of the finite dimensional Lie algebras. In this
regard, notice that as far as these Dynkin diagrams are concerned, those
classical ones are given by infinite series A$_{n}$, B$_{n}$, C$_{n}$, D$%
_{n} $ as depicted by the Figure \textbf{\ref{DD}}. Similar comments can be
said about the exceptional ones; especially the E$_{6}$ and E$_{7}$ which
are relevant for the present study. These diagrams have nodes labeled by the
simple roots $\alpha _{i}$ of G; and links $l_{ij}$ given by their non
trivial intersections $\alpha _{i}.\alpha _{j}$.\
Notice also that formally speaking, the Dynkin diagrams considered here are
graphic representations of the Cartan matrices $K_{ij}\left( G\right) $ of
finite dimensional Lie algebras underlying gauge symmetries of QFT's. These
intersection matrices have particular integer entries with $\det K>0$; it
reads in terms of the simple $\alpha _{i}$'s and their co-roots $\alpha
_{j}^{\vee }$ as follows,
\begin{equation}
K_{ij}=\alpha _{i}^{\vee }.\alpha _{j}\qquad ,\qquad \alpha _{i}^{\vee }=%
\frac{2}{\alpha _{i}.\alpha _{i}}\alpha _{i}.
\end{equation}%
where the scalar product $\alpha _{i}.\alpha _{i}$ refers to the length of
the root. It is equal to 2 for the simply laced ADE Lie algebras; thus
leading to a symmetric matrix $K_{ij}$. This feature does not hold for the
non simply laced BC Lie algebras to be also investigated later. \newline
Returning to the minuscule Lax operators $\mathcal{L}_{G}^{\mu }$; they are
associated to the Dynkin diagrams $K_{ij}\left( G\right) $ whose minuscule
node $\mu $ is omitted. As illustration, we give in the Figure \textbf{\ref%
{LG}} four examples regarding the cutting of the minuscule node in the
Dynkin diagram. The first example is given by the omission of the first node
$\alpha _{1}$ in the Dynkin diagram of sl$\left( 9\right) $. This omission
breaks the diagram $K\left( A_{8}\right) $ into two pieces given by%
\begin{equation}
K\left( A_{8}\right) \rightarrow K\left( A_{1}\right) \oplus K\left(
A_{7}\right)
\end{equation}%
with the isolated $\alpha _{1}$ corresponding to $K\left( A_{1}\right) $ and
the seven others to $K\left( A_{7}\right) $. The second graph in the Figure
\textbf{\ref{LG}} represents the omission of the fourth node $\alpha _{4}$
of the Dynkin diagram of sl$\left( 9\right) $. This cutting leads to the
breaking of $K\left( A_{8}\right) $ into three pieces with graphs as%
\begin{equation}
K\left( A_{8}\right) \rightarrow K\left( A_{3}\right) \oplus K\left(
A_{1}\right) \oplus K\left( A_{4}\right)
\end{equation}%
where the isolated $\alpha _{4}$ corresponds to $K\left( A_{1}\right) $ and
the two other pieces to $K\left( A_{3}\right) $ and $K\left( A_{4}\right) $.
These two ways of cutting the Dynkin diagram\ of $K\left( A_{8}\right) $
describe two different Lax operators $\mathcal{L}_{sl_{9}}^{\mu _{1}}$ and $%
\mathcal{L}_{sl_{9}}^{\mu _{4}}$ in the sl$_{9}$ theory. Regarding the other
two graphic examples in the Figure \textbf{\ref{LG}}, they concern the
orthogonal so$\left( 18\right) $ Lie algebra. The two cuttings correspond to
the following graph decompositions%
\begin{eqnarray}
K\left( D_{9}\right) &\rightarrow &K\left( A_{1}\right) \oplus K\left(
D_{8}\right)  \label{d8} \\
K\left( D_{9}\right) &\rightarrow &K\left( A_{1}\right) \oplus K\left(
A_{8}\right)  \label{a7}
\end{eqnarray}%
The first preserving the orthogonal structure of the broken diagram as shown
by the $K\left( D_{8}\right) $ part. The second cutting destroys this
feature since we have $K\left( A_{8}\right) $. As for the two previous
example, eqs(\ref{d8}-\ref{a7}) describe two different Lax operators in so$%
\left( 18\right) $ theory namely $\mathcal{L}_{so_{18}}^{\mu _{1}}$ and $%
\mathcal{L}_{so_{18}}^{\mu _{9}}$. \
In conclusion, a general classification of the minuscule $\mathcal{L}%
_{G}^{\mu }$-operators describing crossing Wilson and 't Hooft lines with
unit electric and unit magnetic charges is given by the Table\  \textbf{\ref%
{T2}}.
\begin{table}[h]
\centering \renewcommand{\arraystretch}{1.2} $%
\begin{tabular}{|c|c|c|c|c|c|}
\hline
G$_{n}$ & $L_{G_{n}}^{\mu }$ & $R$ of W$_{\xi _{z}}^{R}$ & simple roots & \#
of $L_{G_{n}}^{\mu }$ & $L_{G_{n}}^{\mu }$ -matrix \\ \hline \hline
A$_{n}$ & $L_{A_{n}}^{\mu _{k}}$ & n & $\alpha _{k}$ & \  \ n & eq(\ref{An})
\\ \hline
B$_{n}$ & $L_{B_{n}}^{\mu _{1}}$ & 2n+1 & $\alpha _{1}$ & \  \ 1 & eq(\ref{Bn}%
) \\ \hline
C$_{n}$ & $L_{C_{n}}^{\mu _{n}}$ & 2n & $\alpha _{n}$ & \  \ 1 & eq(\ref{Cn})
\\ \hline
D$_{n}$ & $L_{D_{n}}^{\mu _{1}},L_{D_{n}}^{\mu _{n-1}},L_{D_{n}}^{\mu _{n}}$
& 2n & $\alpha _{1},\alpha _{n-1},\alpha _{n}$ & \  \ 3 & eqs(\ref{Dn1}-\ref%
{Dn2}-\ref{Dn3}) \\ \hline
E$_{6}$ & $L_{E_{6}}^{\mu _{1}},L_{E_{6}}^{\mu _{5}}$ & 27 & $\alpha
_{1},\alpha _{5}$ & \  \ 2 & eqs(\ref{E61}-\ref{E62}) \\ \hline
E$_{7}$ & $L_{E_{7}}^{\mu _{1}}$ & 56 & $\alpha _{6}$ & \  \ 1 & eq(\ref{E7})
\\ \hline
\end{tabular}%
$%
\caption{The full list of the minuscule Lax-operators $L_{G_{n}}^{\protect%
\mu }$ in 4D Chern-Simons theory with gauge symmetry G and crossing Wilson
and 't Hooft lines having unit electric and magnetic charges. This basic
list contains $\left( \mathbf{i}\right) $ four infinite families \ given by
the A$_{N}$-, B$_{N}$- C$_{N}$-, D$_{N}$- types. $\left( \mathbf{ii}\right) $
a finite exceptional subset having three elements: two equivalent E$_{6}$
and one E$_{7}$.}
\label{T2}
\end{table}
In this classification table, we have also given some useful informations
like the electric charge of the Wilson line. We have even anticipated the
results of this study by giving the links to the equations referring to the
explicit expressions of the matrix representations of the Lax operators.
These links to the $\mathcal{L}_{G_{n}}^{\mu }$ expressions are reported in
the last column of the Table\  \textbf{\ref{T2}}. It should be noted here
that these realisations of the $\mathcal{L}_{G_{n}}^{\mu }$'s are derived
with details in section 3. Notice moreover, that some of the results
reported in this table are completely new, in particular those regarding $%
\mathcal{L}_{B}^{\mu }$ and $\mathcal{L}_{C}^{\mu }$ which are further
investigated in sections 4 and 5.

\subsection{Two approaches for constructing the $\mathcal{L}_{G_{n}}^{%
\protect \mu }$'s}

There are two basic approaches to construct the minuscule- Lax matrices of
lower dimensional integrable systems. $\left( 1\right) $ by using Yangian
algebra representations \textrm{\cite{1E}}. $\left( 2\right) $ by following
the 4D Chern-Simons gauge theory approach. To fix the ideas, we review below
the main lines of these two methods.

\subsubsection{Algebraic Yangian approach}

Here, we describe how the Yangian algebra naturally arises in the framework
of the Yang- Baxter equation and its RLL representation whose its solution
leads to the Lax-operator we are interested in. We start by recalling that
the Yangian algebra is an infinite-dimensional Hopf algebra giving a simple
example for quantum groups. Below, we consider the case of homogeneous
rational spin chains with A-type symmetry and we illustrate through a short
way the key steps for building the Lax-matrix $\mathcal{L}_{gl_{n}}\left(
z\right) $ in this setup. We start from the usual Yang Baxter equation (YBE)
given by \textrm{\cite{2A}}
\begin{equation}
R_{12}(z_{1}-z_{2})R_{13}(z_{1}-z_{3})R_{23}(z_{2}-z_{3})=R_{23}(z_{2}-z_{3})R_{13}(z_{1}-z_{3})R_{12}(z_{1}-z_{2})
\end{equation}%
where the R-matrix $R_{ij}(z_{i}-z_{j})$ depends only on the spectral
parameter $z$. This matrix acts on the tensor product of the two vector
spaces $V_{i}\otimes V_{j}$; it describes the coupling of two particles'
worldlines with inner spaces $V_{i}$ and $V_{j}.$ By setting $u=u_{1}-u_{3}$
and $v=u_{2}-u_{3}$ as well as $u-v=u_{1}-u_{2},$ we can bring the above YBE
to the following form where the number of free parameters is reduced down:
(u,v) instead of ($z_{1},z_{2},z_{3}$),%
\begin{equation}
R_{12}(u-v)R_{13}(u)R_{23}(v)=R_{23}(v)R_{13}(u)R_{12}(u-v)  \label{YBE}
\end{equation}%
A quasi-classical solution to this equation is given by $R_{12}(z)=z\left( I+%
\frac{\hbar }{z}P_{12}\right) $ \textrm{\cite{2A}} where P$_{12}$ is the
permutation operator acting like $P_{12}\left( z_{1}\otimes z_{2}\right)
=z_{2}\otimes z_{1}.$ If we take two spaces in (\ref{YBE}) as $V_{1}=\mathbb{%
C}^{n}$ and $V_{2}=\mathbb{C}^{n}$ and leave the third $V_{3}$ unspecified (%
\textrm{auxiliary oscillator space)}, we recover the RLL relations that read
as follows%
\begin{equation}
R_{12}(u-v)L_{13}(u)L_{23}(v)=L_{23}(v)L_{13}(u)R_{12}(u-v)  \label{R}
\end{equation}%
Here, $L_{13}(u)$ and $L_{23}(v)$ are the Lax- matrices we are interested
in; these are n$\times $n matrices constrained by the following commutation
relations obtained after substituting with $R_{12}(z)=z\left( I+\frac{\hbar
}{z}P_{12}\right) .$%
\begin{equation}
\lbrack L_{13}(u),L_{23}(v)]=\frac{\hbar }{u-v}P_{12}\left \{
L_{13}(v)L_{23}(u)-L_{13}(u)L_{23}(v)\right \}  \label{com}
\end{equation}%
To get the above $L_{13}$ and $L_{23},$ we use the so-called monodromy
matrix $M(z)$ with the following properties. It is a n$\times $n matrix
operator function of the complex spectral parameter z that $\left( i\right) $
satisfies the RLL relations (\ref{R}-\ref{com})\ and $\left( ii\right) $\
allows to write down explicit commutation relations defining the Yangian
algebra. By using the canonical matrix generator basis $\left \{ \mathbf{e}%
_{ab}\right \} $\ with 1$\leq a,b\leq $n, we can expand $M(z)$ as follows
\begin{equation}
M(z)=\sum_{a,b=1}^{n}M_{ab}(z)\otimes \mathbf{e}^{ab}  \label{eab}
\end{equation}%
\ with matrix elements $M_{ab}(z)$ analytic in the complex spectral
parameter z. These matrix entries have the following Laurent expansion \cite%
{1F}
\begin{equation}
M_{ab}(z)=M_{ab}^{{\small [0]}}+M_{ab}^{{\small [1]}}z^{-1}+M_{ab}^{{\small %
[2]}}z^{-2}+...  \label{laur}
\end{equation}%
with $M_{ab}^{{\small [r]}}$ being the Laurent modes. Putting the expansion (%
\ref{laur}) into (\ref{com}), we obtain the following system of commutation
relations defining the Yangian algebra generated by the $M_{ab}^{[r]}$'s
\begin{equation}
\lbrack M_{ab}^{{\small [r]}},M_{cd}^{{\small [s]}}]=\sum_{q=1}^{\min
(r,s)}\left( M_{cb}^{{\small [r+s-q]}}M_{ad}^{{\small [q-1]}}-M_{cb}^{%
{\small [q-1]}}M_{ad}^{{\small [r+s-q]}}\right)  \label{mon}
\end{equation}%
with $\min (r,s)$ designating the minimal integer of the pair $\left(
r,s\right) $. \textrm{In what follows}, we briefly describe the construction
of Lax matrices of A-type by using the Yangian algebra Y(gl$_{n}$). This is
an algebraic method that yields the oscillator realisation of the A-type Lax
operator $\mathcal{L}(z)$. We consider the particular case where the
expansion (\ref{laur})\ ends at the first order, the $L(z)$ is therefore
taken as%
\begin{equation}
L(z)=M^{{\small [0]}}+\frac{1}{z}e^{ab}\otimes M_{ba}^{{\small [1]}}
\label{lz}
\end{equation}%
with $M^{{\small [r]}}$ required to satisfy the Yangian algebra. We show
below that the solutions of the RLL\ equations (\ref{mon}) are characterized
by projectors $\Pi _{p}$ and the harmonic oscillator algebra encoded within $%
M_{ba}^{{\small [1]}}$. To construct the $M^{{\small [0]}}$ and $M^{{\small %
[1]}}$ matrices in (\ref{lz}), we use (\ref{mon}) expanded as
\begin{equation}
\begin{tabular}{lll}
$\lbrack M_{ab}^{{\small [0]}},M_{cd}^{{\small [0]}}]$ & $=$ & $0$ \\
$\lbrack M_{ab}^{{\small [0]}},M_{cd}^{{\small [1]}}]$ & $=$ & $0$ \\
$\lbrack M_{ab}^{{\small [1]}},M_{cd}^{{\small [1]}}]$ & $=$ & $M_{cb}^{%
{\small [1]}}M_{ad}^{{\small [0]}}-M_{cb}^{{\small [0]}}M_{ad}^{{\small [1]}%
} $%
\end{tabular}
\label{ge}
\end{equation}%
The two first relations show that $M_{ab}^{{\small [0]}}$ matrix operator is
a central element of the Yangian. So, it can be diagonalised by applying the
automorphism $M_{ab}^{{\small [0]}}\rightarrow \tilde{M}_{ab}^{{\small [0]}%
}=B_{1}M_{ab}^{{\small [0]}}B_{2}$\ where $B_{1}$ and $B_{2}$ are two
invertible $n\times n$\ matrices that act trivially in the quantum space $%
V_{3}$. The diagonal $\tilde{M}_{ab}^{{\small [0]}}$ reads in general like $%
q_{a}\tilde{M}_{aa}$ and eventually can be taken as follows
\begin{equation}
M_{k}^{[0]}=diag(\underset{p}{\underbrace{1,1,...,1}},\underset{(n-p)}{%
\underbrace{0,0,...,0}})  \label{M0}
\end{equation}%
with label belonging $1\leq p\leq n.$ Notice that this choice corresponds to
$M_{p}^{[0]}=\Pi _{p}$ describing projectors $\sum_{a=1}^{p}\left \vert
a\right \rangle \left \langle a\right \vert $. For the special case where $%
p=n$, the matrix $M_{n}^{[0]}$ coincides with the identity $I_{n}.$ To
determine $M^{[1]},$ we have to solve the two last relations of (\ref{ge})
with $M_{p}^{[0]}=\Pi _{p}$. To that purpose, we split the label $1\leq
a,b\leq n$ like pairs $\left( \alpha ,\dot{\alpha}\right) $ and $(\beta ,%
\dot{\beta})$ with $1\leq \alpha ,\beta \leq p$ and $p<\dot{\alpha},\dot{%
\beta}\leq n;$ and then think of $M^{[1]}$ as follows
\begin{equation}
M_{p}^{[1]}=\left(
\begin{array}{cc}
A_{\alpha \beta } & B_{\alpha \dot{\beta}} \\
C_{\dot{\alpha}\beta } & D_{\dot{\alpha}\dot{\beta}}%
\end{array}%
\right)  \label{M1}
\end{equation}%
such that A, B, C, and D are respectively $p\times p,$ $p\times \left(
n-p\right) ,$ $\left( n-p\right) \times p$ and $\left( n-p\right) \times
\left( n-p\right) $ matrices. Next, we substitute (\ref{M0}) and (\ref{M1})
into the Yangian algebra representation (\ref{mon}), we obtain the following
commutation relations constraining (\ref{M1}),%
\begin{equation}
\begin{tabular}{lll}
$\lbrack A_{\alpha \beta },A_{\gamma \delta }]$ & $=$ & $\delta _{\beta
\gamma }A_{\alpha \delta }-\delta _{\alpha \delta }A_{\beta \gamma }$ \\
$\lbrack B_{\alpha \dot{\beta}},B_{\gamma \dot{\delta}}]$ & $=$ & $0$ \\
$\lbrack C_{\dot{\alpha}\beta },C_{\dot{\gamma}\delta }]$ & $=$ & $0$ \\
$\lbrack D_{\dot{\alpha}\dot{\beta}},D_{\dot{\gamma}\dot{\delta}}]$ & $=$ & $%
0$%
\end{tabular}
\label{alg}
\end{equation}%
and%
\begin{equation}
\begin{tabular}{lll}
$\lbrack A_{\alpha \beta },B_{\gamma \dot{\delta}}]$ & $=$ & $-\delta
_{\beta \gamma }B_{\alpha \dot{\delta}},$ \\
$\lbrack A_{\alpha \beta },C_{\dot{\gamma}\delta }]$ & $=$ & $+\delta
_{\alpha \delta }C_{\dot{\gamma}\beta }$ \\
$\lbrack A_{\alpha \beta },D_{\dot{\gamma}\dot{\delta}}]$ & $=$ & $0$%
\end{tabular}
\label{ag}
\end{equation}%
as well as%
\begin{equation}
\lbrack B_{\alpha \dot{\beta}},C_{\dot{\gamma}\delta }]=\delta _{\alpha
\gamma }D_{\dot{\gamma}\dot{\beta}}  \label{BCD}
\end{equation}%
and%
\begin{equation}
\lbrack D_{\dot{\alpha}\dot{\beta}},A_{\gamma \delta }]=[D_{\dot{\alpha}\dot{%
\beta}},B_{\gamma \dot{\delta}}]=[D_{\dot{\alpha}\dot{\beta}},C_{\dot{\gamma}%
\delta }]=0  \label{al}
\end{equation}%
The last relations (\ref{al}) show that the $D_{\dot{\alpha}\dot{\beta}}$'s
are central elements of the monodromy algebra. So, assuming $\det D\neq 0$
we can use the remaining freedom, used in putting $M_{ab}^{{\small [0]}%
}\rightarrow B_{1}M_{ab}^{{\small [0]}}B_{2}$ $=\left( \Pi _{p}\right)
_{ab}, $ to also put the $D_{\dot{\alpha}\dot{\beta}}$- matrix as given by $%
\delta _{\dot{\alpha}\dot{\beta}}.$ By substituting into (\ref{BCD}), we end
up with $[B_{\alpha \dot{\beta}},C_{\dot{\gamma}\delta }]=\delta _{\alpha
\gamma }\delta _{\dot{\gamma}\dot{\beta}}$ that is convenient to rewrite
like
\begin{equation}
\lbrack C_{\dot{\gamma}\delta }^{\prime },B_{\alpha \dot{\beta}}]=\delta
_{\alpha \gamma }\delta _{\dot{\gamma}\dot{\beta}}  \label{qha}
\end{equation}%
where we have set $C_{\dot{\gamma}\delta }^{\prime }=-C_{\dot{\gamma}\delta
} $. In this way, the $B_{\alpha \dot{\beta}}$'s are interpreted as
oscillator creators and $C_{\dot{\gamma}\delta }^{\prime }$ as the
annihilators. Notice that this choice is due to the fact that the algebra (%
\ref{BCD}) must admit a definition of a normalized trace over the oscillator
algebras. This trace is different from the usual algebraic trace over gl(n);
it is needed for the construction of transfer matrices and requires that $%
\det D\neq 0;$ see \cite{1F} for further details. Notice also that the
Yangian algebra given by the commutation relations (\ref{alg}-\ref{al}) is
realised as the direct product%
\begin{equation}
Y\left( gl_{p}\right) \simeq gl(p)\otimes \mathcal{H}^{\otimes p(n-p)}
\end{equation}%
Here, the $gl(p)$ algebra is generated by $A_{\alpha \gamma }$ and $\mathcal{%
H}^{\otimes p(n-p)}$ is the tensor product of $p(n-p)$ copies of the
oscillator algebra (\ref{qha}) generated by the quantum harmonic oscillators
$B_{\alpha \dot{\beta}},C_{\dot{\gamma}\delta }$. The generators $A_{\alpha
\beta }$ solving the constraint relations (\ref{alg}-\ref{ag}) are realized
as follows%
\begin{equation}
A_{\alpha \beta }=\hat{J}_{\alpha \beta }-\sum_{\dot{\beta}=1}^{n-p}\left(
B_{\alpha \dot{\beta}}C_{\dot{\beta}\beta }+\frac{1}{2}\delta _{\alpha \beta
}\right)
\end{equation}%
with $\hat{J}_{\alpha \beta }$ being the transpose of $J_{\alpha \beta }.$
Therefore the Lax matrix constructed within the Yangian approach reads in
terms of these oscillators as%
\begin{equation}
L(z)=\left(
\begin{array}{cc}
z\delta _{\alpha \beta }+\hat{J}_{\alpha \beta }-\dsum \limits_{\dot{\beta}%
=1}^{n-p}\left( B_{\alpha \dot{\beta}}C_{\dot{\beta}\beta }+\frac{1}{2}%
\delta _{\alpha \beta }\right) & B_{\alpha \dot{\beta}} \\
-C_{\dot{\alpha}\beta } & \delta _{\dot{\alpha}\dot{\beta}}%
\end{array}%
\right)
\end{equation}%
This Lax operator serves as an elementary building block for other solutions
via the fusion procedure. They are used to build the Baxter operator and the
transfer matrices using quantum harmonic oscillators; for details see \cite%
{1F} and references therein.

\subsubsection{4D Chern Simons theory}

Here, we describe the main lines of the 4D Chern-Simons gauge theory
invariant under gauge symmetry groups G with Lie algebra g having at least
one minuscule coweight $\mu $. The gauge symmetries considered below are as
those given in the Table \textbf{\ref{T2}}. First, we describe the gauge
field action of the CS theory. Then, we give the RLL\ integrability equation
formulated in terms of the Lax-operators and the R-matrix of Yang-Baxter
equations.

\  \  \  \  \

\textbf{Topological CS gauge field action}:\newline
The 4D Chern-Simons gauge theory is a topological QFT that was first
obtained in \textrm{\cite{1A}}. Its gauge invariant field action $\mathcal{S}%
\left[ \mathcal{A}\right] $ describing the dynamics of the gauge field $%
\mathcal{A}$ in the 4D space, that we take like $\boldsymbol{M}_{4}=\mathbb{R%
}^{2}\times \mathbb{CP}^{1},$ reads as follows
\begin{equation}
\mathcal{S}\left[ \mathcal{A}\right] =\int_{\boldsymbol{M}_{4}}dz\wedge
tr\Omega _{3}  \label{ac}
\end{equation}%
where $\Omega _{3}$\ is the CS 3-form%
\begin{equation}
\Omega _{3}=\mathcal{A}\wedge d\mathcal{A}+\frac{2}{3}\mathcal{A}\wedge
\mathcal{A}\wedge \mathcal{A}
\end{equation}%
To fix the ideas, we consider the family of gauge symmetry $G=SL_{N},$ a
similar description is valid for the gauge symmetries given in the Table \ref%
{T2}. In this case, the 1-form gauge potential is a function of the
variables $\left( x,y;z\right) $ parameterising $\boldsymbol{M}_{4}$. It
expands like $\mathcal{A}=t_{a}\mathcal{A}^{a}$ with $t_{a}$ standing for
the generators of the Lie algebra $sl_{N}$ and $\mathcal{A}^{a}$ a partial
gauge connection as follows \textrm{\cite{1A}}
\begin{equation}
\mathcal{A}^{a}=dx\mathcal{A}_{x}^{a}+dy\mathcal{A}_{y}^{a}+d\bar{z}\mathcal{%
A}_{\bar{z}}^{a}
\end{equation}%
The missing component $dz\mathcal{A}_{z}^{a}$ is killed by the factor $%
dz\wedge $ in the integral measure in (\ref{ac}). The equation of motion of
the potential field $\mathcal{A}$ is given by the vanishing gauge curvature
\begin{equation}
\mathcal{F}_{2}=d\mathcal{A}+\mathcal{A}\wedge \mathcal{A}=0
\end{equation}%
This\ flat curvature property agrees with the topological nature of the CS
theory (\ref{ac}), it indicates that the 4D CS gauge system is in the ground
state with zero energy.

\  \  \  \

\begin{figure}[tbph]
\begin{center}
\includegraphics[width=12cm]{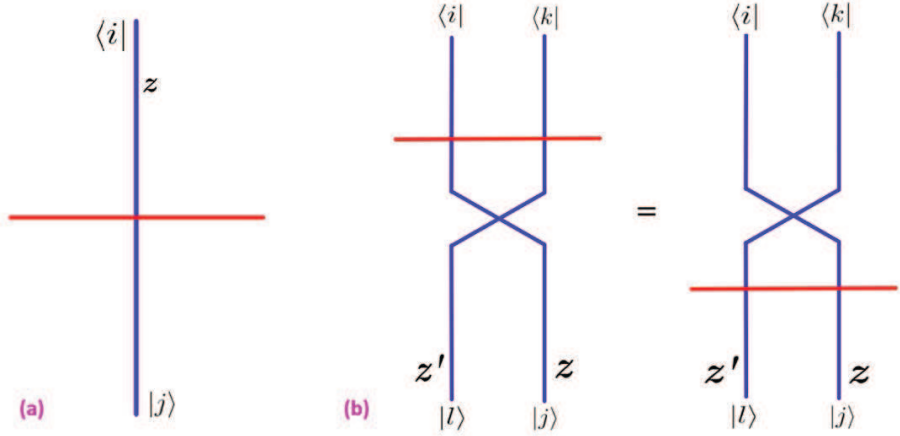}
\end{center}
\par
\vspace{-0.5cm}
\caption{On the left, the operator $\mathcal{L}$ describing the coupling
between a 't Hooft line at z=0 and a Wilson line at z with incoming $%
\left
\langle i\right \vert $ and out going $\left \vert j\right \rangle $
states. On the right the graphic representation of RLL relations.}
\label{Lop}
\end{figure}

\textbf{Line defects and observables}:\newline
To deform the topological ground state, we insert in the 4D CS gauge theory
electrically charged Wilson and magnetically charged 'tHooft lines with
magnetic charge given by the minuscule coweights. Interesting cases
correspond to having interacting Wilson and 'tHooft lines. Examples of such
couplings are given by crossing lines as depicted in the Figures \textbf{\ref%
{Lop}-(a)} and\textbf{\  \ref{Lop}-(b) }having an interpretation in terms of
the Lax-operator and the RLL integrability equation.\
Let us comment the Figure \textbf{\ref{Lop}-(a) }while focussing on SL$_{N}$
theory; we have:

\begin{description}
\item[$\left( \mathbf{1}\right) $] \  \ an electrically charged Wilson W$^{%
\mathbf{R}}$ with quantum states $\left \vert i\right \rangle $ generating a
representation \textbf{R} of the gauge symmetry. This \textbf{R} is taken as
the fundamental representation \textbf{N} of the gauge symmetry SL$_{N}$.
The Wilson observable is given by the vertical blue line in the Figure
\textbf{\ref{Lop}}-(\textbf{a)}.

\item[$\left( \mathbf{2}\right) $] \  \ a magnetically 't Hooft line tH$^{\mu
}$ with magnetic charge given by one of the N-1 minuscule coweights; say the
first coweight $\mu _{1}$ in the Tables \textbf{\ref{T1}}-\textbf{\ref{T2}}.
The magnetic 't Hooft observable is given by the horizontal red line in the
Figure \textbf{\ref{Lop}}-(\textbf{a)}.
\end{description}

\  \  \newline
The crossing matrix $\left \langle W^{\mathbf{R}},tH^{\mu }\right \rangle $
describing the coupling of the two topological lines is given by the Lax
operator $\mathcal{L}_{sl_{N}}^{\mu }$ we are interested in here. It is
obtained by solving the so-called RLL integrability equations briefly
described here below.

\  \  \  \

\subsubsection{RLL\ equation satisfied by the Lax operators}

An interesting way to deal with the properties of the phase space $\mathcal{E%
}_{ph}$ of the classical Lax operator $\mathcal{L}$ is to use the graphic
representation given by the Figures \textbf{\ref{Lop}-(a)} and\textbf{\ (b)}%
. In the left picture of the this figure, we think of $\mathcal{L}%
_{sl_{N}}^{\mu }$ as a matrix operator $L_{i}^{j}=\left \langle i|\mathcal{L}%
_{sl_{N}}^{\mu }|j\right \rangle $ describing the crossing of a horizontal
't Hoof line with a vertical Wilson line on which propagate $\left \langle
i\right \vert $- and $\left \vert j\right \rangle $- states. The symplectic
structure of the phase space of two operators $L_{j}^{r}\left( z\right) $
and $L_{l}^{s}\left( w\right) ,$ with spectral parameters z and w, is given
by the RLL relations shown \textrm{in} the Figure \textbf{\ref{Lop}}. These
integrability relations (which also hold at the quantum level) are due to
the topological invariance of the 4D CS theory and read explicitly as
follows \textrm{\cite{2A},}%
\begin{equation}
R_{rs}^{ik}\left( z-w\right) L_{j}^{r}\left( z\right) L_{l}^{s}\left(
w\right) =L_{r}^{i}\left( w\right) L_{s}^{k}\left( z\right)
R_{jl}^{rs}\left( z-w\right)  \label{rll}
\end{equation}%
Generally speaking, this tensorial relation gives the quantum integrability
conditions of the topological system. The four rank object $%
R_{rs}^{ik}\left( z-w\right) $ is the usual R-operator used in the study of
Yang- Baxter equation. At the leading order in the $\hbar $-expansion, we
have $R_{jl}^{ik}(z)\simeq \delta _{j}^{i}\delta _{l}^{k}+\frac{\hbar }{z}%
c_{jl}^{ik}+O(h%
{{}^2}%
)$ with $c_{jl}^{ik}$\ standing for the double Casimir of the gauge
symmetry. Putting back into \ref{rll}, we obtain the following classical
RLL\ relations
\begin{equation}
\left \{ L_{j}^{i}\left( w\right) ,L_{l}^{k}\left( z\right) \right \} _{PB}=%
\frac{1}{z-w}\left[ L_{l}^{i}\left( z\right) L_{j}^{k}\left( w\right)
-L_{l}^{i}\left( w\right) L_{j}^{k}\left( z\right) \right]
\end{equation}%
where $\left \{ \ast ,\ast \right \} _{PB}$\ stands for the Poisson Bracket.%
\newline
One of the key points about eqs(\ref{rll}) is that they are solved by the
following minuscule Lax operators
\begin{equation}
\mathcal{L}_{sl_{N}}^{\mu }\left( z\right) =e^{X}e^{\mu }e^{Y}  \label{L}
\end{equation}%
This solution first obtained in \textrm{\cite{4D}} concerns gauge symmetries
$G$ having minuscule coweight $\mu $ (here $SL_{N}$). The $X$ and $Y$ are
special matrices having the typical expansions $X=b^{\alpha }X_{\alpha }$
and $Y=Y^{\alpha }c_{\alpha }$ where the $X_{\alpha }$'s and $Y^{\alpha }$'s
are generators of the nilpotent subalgebras $\boldsymbol{n}_{+}$ and $%
\boldsymbol{n}_{-}$ involved in the Levi-decomposition of the Lie algebra of
the gauge symmetry with respect to the minuscule coweight \textrm{\cite{Levi}%
. }This decomposition is given in this case by,
\begin{equation}
\begin{tabular}{lll}
$g$ & $=$ & $\boldsymbol{l}_{\mu }\oplus \boldsymbol{n}_{+}\oplus
\boldsymbol{n}_{-}$ \\
$sl_{N}$ & $=$ & $\left( sl_{1}\oplus sl_{N-1}\right) \oplus \left(
N-1\right) _{+}\oplus \left( N-1\right) _{-}$%
\end{tabular}%
\end{equation}%
Another interesting point concerning (\ref{rll}) is that at the leading
order in the $\hbar $-expansion of the R- matrix, one finds that the $\left(
b,c\right) $ parameters involved in the expansions $b^{\alpha }X_{\alpha }$
and $Y^{\alpha }c_{\alpha }$ satisfy the usual Poisson bracket $\left \{
b^{\alpha },c_{\beta }\right \} _{PB}=\delta _{\beta }^{\alpha }$ of the
symplectic geometry. Then, this classical limit teaches us that: $\left(
\mathbf{a}\right) $ the $\left( b^{\alpha },c_{\alpha }\right) $ are nothing
but the Darboux coordinates (classical harmonic oscillators) of the phase
space of the L-operator. $\left( \mathbf{b}\right) $ the particular solution
(\ref{L}) gives an oscillator realisation of the Lax operators like the ones
obtained from the Yangian based method of integrable quantum spin chains.

\section{Minuscule $\mathcal{L}_{G}$ from CS theory: the full list}

In this section, we introduce and develop a suitable operator basis (method
of projectors) to deal with the explicit calculation of the Lax operator of
integrable spin chain satisfying the RLL equation. We begin by describing
the main steps of the derivation of the minuscule Lax operators\textrm{\ }$%
\mathcal{L}_{G}$\textrm{\ }for gauge symmetries G given by the A$_{N}$, B$%
_{N}$, C$_{N}$, D$_{N},$ E$_{6}$, E$_{7}$\ families. Then, we give the full
list of the explicit oscillator realisations of these $\mathcal{L}_{G}$s.
This list\ gives a unified description of all minuscule $\mathcal{L}_{G}$s
and presents new results concerning the non simply laced families\textrm{\ }$%
\mathcal{L}_{B}^{\mu _{1}}$\textrm{\ }and\textrm{\ }$\mathcal{L}_{C}^{\mu
_{N}}$\textrm{. }Other aspects regarding the relationships between the
various\textrm{\ }$\mathcal{L}_{G}$'s and discrete symmetries are also
studied in order to complete the investigation of minuscule $\mathcal{L}_{G}$%
s.

\subsection{The calculation of $\mathcal{L}_{G}$: method of projectors}

Here we give a suitable calculation method to work out the explicit
matrix-representation of the minuscule Lax operator $\mathcal{L}_{G}$. We
term this approach as the method of projectors; for the motivation behind
this terminology, we refer to eqs(\ref{56}-\ref{rde}) given below. \newline
We start by recalling that the minuscule Lax operator $\mathcal{L}_{G}$ is
given by (\ref{L}) namely $\mathcal{L}_{G}^{\mu }=e^{X}z^{\mu }e^{Y}$. This
formula involves three Lie algebraic objects that we comment below:

\begin{itemize}
\item The adjoint action $\mathbf{\mu }$ of the minuscule coweight; it
belongs to the Lie sub-algebra $\boldsymbol{l}_{\mu }$ of Table \textbf{\ref%
{T1}}. This $\boldsymbol{l}_{\mu }$ results from the Levi- decomposition of
the Lie algebra $g$ underlying the gauge symmetry which decomposes like $g=%
\boldsymbol{n}_{-}\oplus \boldsymbol{l}_{\mu }\oplus \boldsymbol{n}_{+}$.

\item The two $X$ and $Y$ operators are nilpotent matrices evaluated in the
nilpotent sub-algebras $\boldsymbol{n}_{+}$ and $\boldsymbol{n}_{-}$
following the Levi- decomposition. The intrinsic properties of $X$ and $Y$
carry data on the electric charge of the Wilson line and the magnetic charge
of the 't Hooft line.
\end{itemize}

\  \  \  \newline
To determine the explicit expression of $\mathcal{L}_{G}$, we have to first
find the matrix representations of these three objects; then use them to
calculate $e^{X}z^{\mathbf{\mu }}e^{Y}$. \newline
So, given a 4D Chern-Simons theory with gauge symmetry G, we can determine
the expression of the minuscule $\mathcal{L}_{G}$ solving the RLL equation (%
\ref{rll}) by using (\ref{L}). This is done in three steps as described
below:

\  \  \newline
$\mathbf{1)}$ \textbf{Working out the adjoint action of}\emph{\ }$\mu $%
\newline
The adjoint action of the minuscule coweight is a charge operator
representation $\mathbf{\mu }$ acting on the Hilbert space $\mathcal{H}$ of
the Wilson line interacting with the 't Hooft defect. For concreteness, we
denote the vector basis of the space $\mathcal{H}$ of the quantum states by
kets $\left \vert i\right \rangle $ as in the Figure \ref{Lop}-(\textbf{a}).
These quantum states generating the representation $\mathbf{R}_{G}$ of the
gauge symmetry run along the Wilson line. This representation $\mathbf{R}%
_{G} $ will be thought of below as given by the fundamental representation
of the gauge symmetry.\newline
Because of the Levi-decomposition of the gauge symmetry G, the
representation $\mathbf{R}_{G}$ splits in turns into a sum of
representations $\mathbf{R}_{a}^{\boldsymbol{l}_{\mu }}$ of the
Levi-subalgebra $\boldsymbol{l}_{\mu }$. Formally, we have
\begin{equation}
\mathbf{R}_{G}=\oplus _{a}\mathbf{R}_{a}^{\boldsymbol{l}_{\mu }}  \label{56}
\end{equation}%
The charge operator $\mathbf{\mu }$ is effectively given by these
representations; it explicitly reads in terms of projectors $\varrho _{_{%
\mathbf{R}_{a}}}:\mathbf{R}_{G}\rightarrow \mathbf{R}_{a}$ as follows
\begin{equation}
\mathbf{\mu }=\sum_{\mathbf{R}_{a}}m_{_{\mathbf{R}_{a}}}\varrho _{_{\mathbf{R%
}_{a}}}  \label{fge}
\end{equation}%
with%
\begin{equation}
\sum_{\mathbf{R}_{a}}\varrho _{_{\mathbf{R}_{a}}}=I_{id}\qquad ,\qquad
\varrho _{_{\mathbf{R}_{a}}}\varrho _{_{\mathbf{R}_{b}}}=\delta _{ab}\varrho
_{_{\mathbf{R}_{a}}}
\end{equation}%
For a traceless representation $\mathbf{R}_{G},$ we get the following
constraint relation on the Levi-charges $m_{_{\mathbf{R}_{a}}}$ carried by a
representation $\mathbf{R}_{a}^{\boldsymbol{l}_{\mu }}$.
\begin{equation}
\sum_{\mathbf{R}_{a}}m_{_{\mathbf{R}_{a}}}=0  \label{rde}
\end{equation}%
$\mathbf{2)}$ \textbf{Solving the conditions of the Levi-decomposition}%
\newline
Once the adjoint action $\mathbf{\mu }$ is known, we move to determining the
realisation of the $X$ and $Y$ matrices in the Hilbert space of quantum
states of the Wilson/'t Hooft lines. They are obtained by $\left( i\right) $
using the expansions $X=b^{\alpha }X_{\alpha }$ and $Y=c_{\alpha }Y^{\alpha
} $; and $\left( ii\right) $ solving the Levi-conditions
\begin{equation}
\begin{tabular}{lll}
$\left[ \mathbf{\mu },X_{\alpha }\right] $ & $=$ & $+X_{\alpha }$ \\
$\left[ \mathbf{\mu },Y^{\alpha }\right] $ & $=$ & $-Y^{\alpha }$ \\
$\left[ X_{\alpha },Y^{\alpha }\right] $ & $=$ & $\delta _{\alpha }^{\alpha }%
\mathbf{\mu }$%
\end{tabular}
\label{lec}
\end{equation}%
In these Levi-constraint relations, Each generator $X_{\alpha }$ of $%
\boldsymbol{n}_{+}$ carries a positive unit charge $+1$ and each generator $%
Y^{\alpha }$ of $\boldsymbol{n}_{-}$\ carries a negative unit charge $-1.$

\  \  \  \newline
$\mathbf{3)}$ \textbf{Calculating the L-operator }$\mathcal{L}_{G}$\textbf{\
as a bi-polynom in }$\mathbf{X}$\textbf{\ and }$\mathbf{Y}$\newline
Because of the nilpotency property of the $X_{\alpha }$ and $Y^{\alpha }$
generators of $\boldsymbol{n}_{\pm }$ and commuting properties like $\left[
X_{\alpha },X_{\beta }\right] =0$; the $X$ and $Y$ are as well nilpotent.
So, there should exist an order n and an order m such that $X^{k+1}=0$ and $%
Y^{l+1}=0.$ In fact the two orders are equal $k=l$ because of the duality
between $X$ and $Y$. So, the exponentials $e^{X}$ and $e^{Y}$ have finite
expansions of the form $e^{Z}=I+Z+...\frac{1}{k!}Z^{k}.$ Putting this back
into $e^{X}z^{\mu }e^{Y},$ we end up with the oscillator realisation of $%
\mathcal{L}_{G}$ given by
\begin{equation}
\mathcal{L}_{G}=\sum_{n,m}\frac{1}{n!m!}X^{n}z^{\mu }Y^{m}
\end{equation}%
where $X=b^{\alpha }X_{\alpha }$ and $Y=c_{\beta }Y^{\beta }$ as well as
\begin{equation}
z^{\mu }=\sum z^{m_{a}}\varrho _{_{\mathbf{R}_{a}}}
\end{equation}%
Finally, using the properties of the projectors $\varrho _{_{\mathbf{R}%
_{a}}} $, we can present $\mathcal{L}_{G}$ by a matrix $L_{ab}$ given by%
\begin{equation}
L_{ab}^{G}=\varrho _{_{\mathbf{R}_{a}}}\mathcal{L}_{G}\varrho _{_{\mathbf{R}%
_{b}}}  \label{ab}
\end{equation}

\subsection{The A$_{N}$\ operator family: $L_{ab}^{A_{N}}$}

Here, we give an example to explain the above steps of the calculation of
the Lax-matrix $L_{ab}^{G}$ (\ref{ab}). We consider the case of a 4D
Chern-Simons theory with gauge symmetry SL$\left( N\right) $ in presence of
an electrically charged Wilson line W with fundamental electric charge $%
\lambda _{1}$ as depicted by the Figure \textbf{\ref{Lop}}-(\textbf{a}).
this W-line crosses a magnetically charged 't Hooft line with magnetic
charge given by the minuscule coweight $\mu _{1}$. \newline
As far as the Lie algebra $A_{N-1}\simeq sl_{N}$ of the gauge symmetry SL$%
\left( N\right) $ is concerned, it is interesting to recall some useful
mathematical features that we collect in the following table (\ref{T3}).
\begin{equation}
\begin{tabular}{|c|c|c|c|c|c|}
\hline
{\small algebra} & ${\small A}_{N-1}$ & ${\small sl}_{1}$ & ${\small A}%
_{N-2} $ & ${\small n}_{+}$ & ${\small n}_{-}$ \\ \hline
{\small dim} & ${\small (N-1})({\small N+1})$ & ${\small 1}$ & ${\small N}%
^{2}{\small -2N}$ & ${\small N-1}$ & ${\small N-1}$ \\ \hline
{\small rank} & ${\small N-1}$ & ${\small 1}$ & ${\small N-2}$ & ${\small 0}$
& ${\small 0}$ \\ \hline
{\small roots} & ${\small N(N-1)}$ & ${\small 0}$ & ${\small (N-1)(N-2)}$ & $%
{\small N-1}$ & ${\small N-1}$ \\ \hline
{\small Cartan H}$_{i}$ & ${\small N-1}$ & ${\small 1}$ & ${\small N-2}$ & $%
{\small 0}$ & ${\small 0}$ \\ \hline
{\small step }$\text{E}_{{\small \pm }\alpha }$ & ${\small 2\times }\frac{%
N(N-1)}{2}$ & ${\small 0}$ & ${\small 2\times }\frac{(N-1)(N-2)}{2}$ & $%
{\small (N-1)}\ {\small X}_{\beta }$ & ${\small (N-1)}\ {\small Y}^{\alpha }$
\\ \hline
\end{tabular}
\label{T3}
\end{equation}%
\begin{equation*}
\text{ \  \  \  \ }
\end{equation*}%
These properties regard the Levi-decomposition of $sl_{N}$ with respect to
the minuscule coweight $\mu _{1}$ namely%
\begin{equation}
sl_{N}\rightarrow sl_{1}\oplus sl_{N-1}+\boldsymbol{n}_{+}\oplus \boldsymbol{%
n}_{-}
\end{equation}%
The adjoint action of the minuscule coweights $\mu _{1}$ is therefore given
by
\begin{equation}
\mathbf{\mu }_{1}=\frac{N-1}{N}\left \vert 1\right \rangle \left \langle
1\right \vert -\frac{1}{N}\sum_{i=2}^{N}\left \vert i\right \rangle \left
\langle i\right \vert
\end{equation}%
The generators $X_{i}$ and $Y^{i}$ of the nilpotent sub-algebras $%
\boldsymbol{n}_{+}$ and $\boldsymbol{n}_{-}$ solving the Levi- constraint
relations are given by
\begin{equation}
\begin{tabular}{lll}
$X_{i}$ & $=$ & $\left \vert 1\right \rangle \left \langle 1+i\right \vert $
\\
$Y^{i}$ & $=$ & $\left \vert 1+i\right \rangle \left \langle 1\right \vert $%
\end{tabular}%
\end{equation}%
The Lax matrix representing the L-operator reads as follows
\begin{equation}
\mathcal{L}_{A_{N}}=\left(
\begin{array}{cc}
z^{\frac{N-1}{N}}+z^{-\frac{1}{N}}\mathbf{b}^{T}\mathbf{c} & z^{-\frac{1}{N}}%
\mathbf{b}^{T} \\
z^{-\frac{1}{N}}\mathbf{c} & z^{-\frac{1}{N}}I_{N-1}%
\end{array}%
\right)
\end{equation}%
with $\mathbf{b}^{T}=\left( b_{1},...,b_{N-1}\right) $ and $\mathbf{c}%
=\left( c_{1},...,c_{N-1}\right) ^{T}.$ By multiplying this matrix with the
factor $z^{\frac{1}{N}}$, we get the familiar Lax-matrix obtained in the
quantum spin chain literature \cite{2D}, namely%
\begin{equation}
\mathcal{\tilde{L}}_{A_{N}}=\left(
\begin{array}{cc}
z^{N-1}+\mathbf{b}^{T}\mathbf{c} & \mathbf{b}^{T} \\
\mathbf{c} & I_{N-1}%
\end{array}%
\right)
\end{equation}

\subsection{The full list of minuscule L-operators}

The list of the full set $\mathfrak{M}$ of minuscule Lax operators with unit
charges is infinite. It contains five sub-families given by the four
classical series A$_{N}$, B$_{N}$, C$_{N}$, D$_{N}$ and the exceptional
finite E$_{6},$ E$_{7}$. These sub-families o the set $\mathfrak{M}$ of
L-operators are revisited below separately in the given ordering.

\subsubsection{A$_{N-1}$- type operators $\mathcal{L}_{A}$}

First we give the general form of the Lax-matrices $\mathcal{L}%
_{A_{N-1}}^{\mu _{k}}$ for generic fundamentals coweights $\mu _{k}$ with
label $1\leq k\leq N-1$ and label $N-1$ being the rank of $A_{N-1}$. Then,
we comment the exotic property of the $A_{N-1}$-family regarding the CS
theory with gauge symmetry SL$_{2M};$ that is $N=2M$.

\  \  \

\textbf{I}) \textbf{Generic formula for} $\mathcal{L}_{A_{N-1}}^{\mu _{k}}$%
\newline
We start by recalling that the A$_{N-1}$ Lie algebra series has $N-1$
minuscule coweights $\mathbf{\mu }_{k}$ labeled by $1\leq k\leq N-1.$ These
coweights are expressed in terms of the weight vector basis $\left \{
e_{i}\right \} _{1\leq i\leq N}$ like
\begin{equation}
(\text{ }\underbrace{\frac{N-k}{N},...,\frac{N-k}{N}}_{k}\text{ };\text{ }%
\underbrace{-\frac{k}{N},...,-\frac{k}{N}}_{N-k}\text{ })
\end{equation}%
The unit vectors $e_{i}$ in the weight vector basis are now on denoted by
the kets $\left \vert i\right \rangle $ and their duals are represented by
the bras $\left \langle i\right \vert $. The Levi-decomposition of the sl$%
\left( N\right) $ Lie algebra with respect to $\mathbf{\mu }_{k}$ is given by%
\begin{equation}
sl_{N}\rightarrow sl_{1}\oplus sl_{k}\oplus sl_{N-k}\oplus \boldsymbol{n}%
_{+}\oplus \boldsymbol{n}_{-}
\end{equation}%
with $\boldsymbol{n}_{\pm }=k\left( N-k\right) _{\pm }$. The adjoint action
of the minuscule coweight is realised in terms of the $\left \vert
i\right
\rangle $ weight vectors as%
\begin{equation}
\mathbf{\mu }_{k}=\left( 1-\frac{k}{N}\right) \dsum \limits_{i=1}^{k}\left
\vert i\right \rangle \left \langle i\right \vert -\frac{k}{N}\dsum
\limits_{i=k+1}^{N}\left \vert i\right \rangle \left \langle i\right \vert
\end{equation}%
For later use, we will also use the convenient notation%
\begin{equation}
\mathbf{\mu }_{k}=\left( 1-\frac{k}{N}\right) \dsum \limits_{i=1}^{k}\left
\vert i\right \rangle \left \langle i\right \vert -\frac{k}{N}\dsum
\limits_{i=1}^{N-k}\left \vert \bar{\imath}\right \rangle \left \langle \bar{%
\imath}\right \vert  \label{a1}
\end{equation}%
where we have set $\bar{\imath}=N+1-i.$ Notice that $i+\bar{\imath}=N+1$
such that%
\begin{equation}
\begin{tabular}{c|c|c|c|c|c}
$i$ & $1$ & $2$ & $\cdots $ & $N-1$ & $N$ \\ \hline
$\bar{\imath}$ & $\bar{1}=N$ & $\bar{2}=N-1$ & $\cdots $ & $\overline{N-1}=2$
& $\bar{N}=1$%
\end{tabular}%
\end{equation}%
The charge operator (\ref{a1}) is traceless, $Tr\left( \mathbf{\mu }%
_{k}\right) =0$. We often refer to $\mathbf{\mu }_{k}$ as the Levi-Charge
operator. The $k\left( N-k\right) $ operators generating the nilpotent
algebras $\boldsymbol{n}_{\pm }$ are now denoted like $X_{ia}^{\mu _{k}}$
and $Y_{\mu _{k}}^{ia}.$ The explicit representation of these generators in
the Hilbert space of the crossing lines is obtained by solving the
Levi-constraint relations (\ref{lec}) reading as follows%
\begin{equation}
\begin{tabular}{lll}
$X_{i\bar{a}}^{\mu _{k}}$ & $=$ & $\left \vert i\right \rangle \left \langle
\bar{a}\right \vert $ \\
$Y_{\mu _{k}}^{\bar{a}i}$ & $=$ & $\left \vert \bar{a}\right \rangle \left
\langle i\right \vert $%
\end{tabular}
\label{a2}
\end{equation}%
with labels as $1\leq i\leq k$ and $k<a\leq N$ or equivalently $\bar{N}\leq
\bar{a}<\bar{k}$. The Lax- matrix associated with the coweight $\mu _{k}$ is
obtained by substituting the expansions $X=b^{i\bar{a}}X_{i\bar{a}}^{\mu
_{k}}$ and $X=Y^{\bar{a}i}c_{\bar{a}i}$ into $e^{X}z^{\mathbf{\mu }%
_{k}}e^{Y} $. We get
\begin{equation}
\mathcal{L}_{A_{N}}^{\mu _{k}}=\left(
\begin{array}{cc}
z^{\frac{N-k}{N}}I_{k}+z^{-\frac{k}{N}}\mathbf{bc} & z^{-\frac{k}{N}}\mathbf{%
b} \\
z^{-\frac{k}{N}}\mathbf{c} & z^{-\frac{k}{N}}I_{N-k}%
\end{array}%
\right)  \label{An}
\end{equation}%
Here, the $\mathbf{b}$ and $\mathbf{c}$ refer to the k$\left( N-k\right) $
oscillators $\mathbf{b}^{i\bar{a}}$ and k$\left( N-k\right) $ oscillators $%
\mathbf{c}_{\bar{a}i}$ given by
\begin{equation}
\mathbf{b}^{ia}=\left(
\begin{array}{ccc}
b^{1\bar{N}} & \cdots & b^{1\overline{k+1}} \\
\vdots & \ddots & \vdots \\
b^{k\bar{N}} & \cdots & b^{k\overline{k+1}}%
\end{array}%
\right) \qquad ,\qquad \mathbf{c}_{ai}=\left(
\begin{array}{ccc}
c_{\bar{N}k} & \cdots & c_{\bar{N}1} \\
\vdots & \ddots & \vdots \\
c_{\overline{k+1}k} & \cdots & c_{\overline{k+1}1}%
\end{array}%
\right)  \label{An1}
\end{equation}

\textbf{II}) \textbf{Automorphism symmetries of} $\mathcal{L}_{A_{N-1}}^{\mu
_{k}}\newline
$We begin by noticing that the examination of eq(\ref{a1}) reveals a
remarkable property of the coweights $\mathbf{\mu }_{k}$. In fact, by making
the following $\mathbb{Z}_{2}$ discrete change,
\begin{equation}
\mathbb{Z}_{2}:\left.
\begin{array}{ccc}
k & \rightarrow & N-k \\
\left \vert i\right \rangle & \rightarrow & \left \vert \bar{\imath}\right
\rangle%
\end{array}%
\right.  \label{z2}
\end{equation}%
we learn that the corresponding operator charge operators transform as $%
\mathbf{\mu }_{N-k}\rightarrow -\mathbf{\mu }_{k}.$ This symmetry property
indicates that eq(\ref{An}) hides an interesting symmetry property
explicitly exhibited by the case where $N=2M$. This symmetry property has an
interpretation in the language of $A_{N-1}$ representations and their
adjoint conjugates. It also allows to engineer new $\mathcal{L}$-operators
from (\ref{An}) inspired by Dynkin diagram folding ideas. We will see later
that the fixed point of the folding of (\ref{An}) under the above $\mathbb{Z}%
_{2}$ symmetry reading as%
\begin{equation}
\mathcal{L}_{A_{2M-1}}^{\mu _{M}}=\left(
\begin{array}{cc}
z^{\frac{1}{2}}I_{M}+z^{-\frac{1}{2}}\mathbf{bc} & z^{-\frac{1}{2}}\mathbf{b}
\\
z^{-\frac{1}{2}}\mathbf{c} & z^{-\frac{1}{2}}I_{M}%
\end{array}%
\right)  \label{a2m}
\end{equation}%
gives precisely the L-operators $\mathcal{L}_{C_{M}}^{\mu }$ of the
symplectic family $SP\left( 2M\right) $. In this regard, notice the\
following properties:

\begin{description}
\item[$\left( \mathbf{1}\right) $ ] \ For the 4D Chern-Simons theory with
gauge symmetry $G=SL\left( 2M\right) ,$ the Lax- matrix $\mathcal{L}%
_{A_{2M-1}}^{\mu _{k}}$ has four main blocks: $\left( i\right) $ two
diagonal blocks of dimensions k$\times $k and (2M-k)$\times $(2M-k) and $%
\left( ii\right) $ two off diagonal with dimensions k$\times $(2M-k) and
(2M-k)$\times $k. For the particular case where $k=M$, all the four above
blocks are iso-dimensional. This iso-dimensional property corresponding to
the fixed point of the $\mathbb{Z}_{2}$-symmetry (\ref{z2}) is nothing but
the $\mathbb{Z}_{2}$-outer-automorphism of the Dynkin diagram of SL$\left(
2M\right) $ as depicted by the Figure \textbf{\ref{sl4} }for the example of
SL$\left( 4\right) $.
\begin{figure}[tbph]
\begin{center}
\includegraphics[width=14cm]{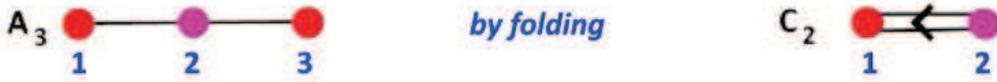}
\end{center}
\par
\vspace{-0.5cm}
\caption{On the left, the Dynkin diagram of the Lie algebra A$_{3}.$ It has
a $\mathbb{Z}_{2}$- outer-automorphism symmetry leaving one node fixed (in
magenta color). On the right\textbf{, }the\textbf{\ }Dynkin diagram of the
symplectic Lie algebra C$_{2}$. It is obtained by folding A$_{3}$ under $%
\mathbb{Z}_{2}$.}
\label{sl4}
\end{figure}

\item[$\left( \mathbf{2}\right) $ ] $\ $For the sub-family of the 4D CS
gauge symmetries given by $SL\left( 2M\right) $, the Levi-decomposition of
its underlying Lie algebra reads as follows
\begin{equation}
sl_{2M}\rightarrow sl_{1}\oplus sl_{M}\oplus sl_{M}\oplus M_{+}^{2}\oplus
M_{-}^{2}  \label{sl2m}
\end{equation}%
This can be checked from the calculation of the ranks and the dimensions of
both sides of (\ref{sl2m}). If thinking of this gauge symmetry in terms of
the semi-simple $gl_{2M}$, its 4M$^{2}$ dimensions split like $%
M^{2}+M^{2}+M^{2}+M^{2}.$ For the special coweight $\mu _{k}$ with $k=M$,
eqs(\ref{a1}-\ref{a2}) read as follows%
\begin{equation}
\mathbf{\mu }_{M}=\frac{1}{2}\dsum \limits_{i=1}^{M}\left( \left \vert
i\right \rangle \left \langle i\right \vert -\left \vert \bar{\imath}\right
\rangle \left \langle \bar{\imath}\right \vert \right)
\end{equation}%
and%
\begin{equation}
\begin{tabular}{lll}
$X_{i\bar{j}}$ & $=$ & $\left \vert i\right \rangle \left \langle \bar{j}%
\right \vert $ \\
$Y^{\bar{\imath}j}$ & $=$ & $\left \vert \bar{\imath}\right \rangle \left
\langle j\right \vert $%
\end{tabular}%
\end{equation}%
The Lax-matrix (\ref{An}) associated with the coweights $\mathbf{\mu }_{M}$
is given by (\ref{a2m}). We will reconsider these relations when we study
the construction of the Lax-matrix for 4D Chern-Simons theory with symplectic%
$SP_{2M}$ gauge symmetry (\textrm{see section 5}).
\end{description}

\subsubsection{B-type operators $\mathcal{L}_{B}$: new result}

\begin{figure}[tbph]
\begin{center}
\includegraphics[width=6cm]{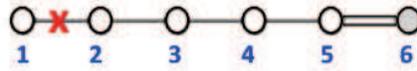}
\end{center}
\par
\vspace{-0.5cm}
\caption{Dynkin Diagram of B$_{N}$ illustrated for the example where $N=6$.
By cutting the first node $\protect \alpha _{1},$ one ends with the Diagram
associated with the Dynkin diagram of the B$_{N-1}$ subalgebra and a free
node $\protect \alpha _{1}$ capturing data on the SO$_{2}$ Levi-charge and
the nilpotent subalgebras.}
\label{B6}
\end{figure}

Now, we consider the case where the gauge symmetry of the 4D Chern-Simons is
given by SO$_{2N+1}$. As shown in Table \textbf{\ref{T1}}, this gauge
symmetry group has one minuscule coweight $\mathbf{\mu }_{1}$ dual to the
simple root $\alpha _{1}$. To get more insight into the algebraic properties
of $\mathcal{L}_{B},$ we recall some useful properties of the $B_{N}$ gauge
symmetry. The SO$_{2N+1}$ has $2N^{2}$ roots, half of them are positive and
denoted as $+\alpha _{ij}^{\pm }$ where $i,j=1,...,N$. The negative ones are
given by the opposites that read as $-\alpha _{ij}^{\pm }$. These $\pm
\alpha _{ij}^{\pm }$'s have two lengths: $N\left( 2N-1\right) $ of them have
length 2 realised in terms of $N$ weight vector basis $\left \{
e_{i}\right
\} $ like $\pm \left( e_{i}\pm e_{j}\right) $ with $i\neq j.$
The \textrm{remaining} $N$ others have length 1; they are given by $\pm
\alpha _{ii}^{+}$ and are realised as $\pm e_{i}.$ The $N$ simple roots $%
\alpha _{i}$ of the Lie algebra of SO$_{2N+1}$ are given by: $\left( \mathbf{%
a}\right) $ $\alpha _{i}=e_{i}-e_{i+1}$ for $i=1,...,N-1$ having\ length%
\textrm{\ 2}. $\left( \mathbf{b}\right) $ $\alpha _{N}=e_{N}$ having length
1. In this basis, the minuscule coweight is given by $\mu _{1}=e_{1}$
obeying $\mu _{1}.\alpha _{1}=1.$ \newline
The Levi-decomposition of the Lie algebra of $SO_{2N+1}$ reads as $%
so_{2}\oplus so_{2N-1}\oplus \boldsymbol{n}_{\pm }$ with nilpotent sector as
$\boldsymbol{n}_{\pm }=(2N-1)_{\pm }$. It corresponds to cutting the first
node $\alpha _{1}$ in the Dynkin diagram of $B_{N}\simeq so_{2N+1}$ as
depicted by the Figure \textbf{\ref{B6}}.

The adjoint action of the minuscule coweight $\mathbf{\mu }$ is given by%
\begin{equation}
\mathbf{\mu }=\frac{1}{2}\varrho _{+}+q\dsum \limits_{i=1}^{2N-1}\varrho
_{i}-\frac{1}{2}\varrho _{-}
\end{equation}%
where we have for commodity inserted the central terms although $q=0;$ this
is because it contributes to $z^{\mathbf{\mu }}.$ The $\varrho _{\pm
}=\left
\vert \pm \right \rangle \left \langle \pm \right \vert $ and the $%
\varrho _{i}=\left \vert i\right \rangle \left \langle i\right \vert $ are
projectors. The matrix realisation of the $X_{i}$ and $Y^{i}$ generators of
the nilpotent algebras are as follows
\begin{equation}
\begin{tabular}{lll}
$X_{i}$ & $=$ & $\left \vert +\right \rangle \left \langle i\right \vert
-\left \vert i\right \rangle \left \langle -\right \vert $ \\
$Y^{i}$ & $=$ & $\left \vert i\right \rangle \left \langle +\right \vert
-\left \vert -\right \rangle \left \langle i\right \vert $%
\end{tabular}%
\end{equation}%
We also have the expansions $X=b^{i}X_{i}$ and $Y=c_{i}Y^{i}$. These
relations will be discussed further in \textrm{section 4}. The explicit
matrix realisation of the Lax operator $\mathcal{L}_{B}^{\mu _{1}}$ is also
derived in section 4; it reads as follows%
\begin{equation}
\mathcal{L}_{B}^{\mu _{1}}=\left(
\begin{array}{lll}
z+\mathbf{b}^{T}\mathbf{c+}\frac{1}{4}z^{-1}\mathbf{b}^{2}\mathbf{c}^{2} &
\mathbf{b}^{T}-\frac{1}{2}z^{-1}\mathbf{b}^{2}\mathbf{c}^{T} & \frac{1}{2}%
z^{-1}\mathbf{b}^{2} \\
\mathbf{c-}\frac{1}{2}z^{-1}\mathbf{c}^{2}\mathbf{b} & I_{2N-1}+z^{-1}%
\mathbf{b}^{T}\mathbf{c} & -z^{-1}\mathbf{b} \\
\frac{1}{2}z^{-1}\mathbf{c}^{2} & -z^{-1}\mathbf{c}^{T} & z^{-1}%
\end{array}%
\right)  \label{Bn}
\end{equation}%
with $\mathbf{b}^{2}=b^{i}\delta _{ij}b^{j}$ and $\mathbf{c}^{2}=c_{i}\delta
^{ij}c_{j}$ as well as $\mathbf{b}^{T}\mathbf{.c}=b^{i}c_{i}$. In the basis $%
\left \{ \left \vert +\right \rangle ,\left \vert i\right \rangle
,\left
\vert -\right \rangle \right \} ,$ the entries of the $X$ and $Y$
are respectively given by $(2N+1)\times 1$ and $1\times (2N+1)$ matrix
oscillators given by $\mathbf{b}^{T}=\left( 0,b_{1},\cdots
,b_{2N-1},0\right) $ and $\mathbf{c}^{T}=\left( 0,c_{1},\cdots
,c_{2N-1},0\right) .$

\subsubsection{C-type operators $\mathcal{L}_{C}$: new result}

\begin{figure}[tbph]
\begin{center}
\includegraphics[width=6cm]{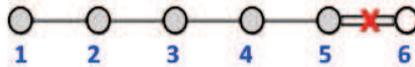}
\end{center}
\par
\vspace{-0.5cm}
\caption{Dynkin Diagram of C$_{N}$ illustrated for the example where $N=6$.
By cutting the first node $\protect \alpha _{N},$ one ends with the Diagram
associated with the Dynkin diagram of the A$_{N-1}$ subalgebra and a free
node $\protect \alpha _{N}$ capturing data on the SO$_{2}$ Levi-charge and
the nilpotent subalgebras.}
\label{C6}
\end{figure}

In this case, the gauge symmetry of the 4D Chern-Simons is given by the
symplectic group SP$_{2N}$ having rank N and dimension $2N^{2}+N$ thought of
below as $N^{2}+N(N+1).$ It has one minuscule coweight $\mu _{N}$ dual to
the simple root $\alpha _{N}$. In this regard, recall that $SP_{2N}$ has $%
2N^{2}$ roots, half of them are positive and the others are negative roots.
These roots are realised in terms of the $e_{i}$ weight vectors as $\pm
\alpha _{ij}^{\pm }=\pm e_{i}\pm e_{j}$ $(1\leq i<j\leq N)$ and $\pm 2e_{i}$
for $1\leq i\leq N.$ The $N$ simple roots $\alpha _{i}$ of SP$_{2N}$ are
given by $e_{i}-e_{i+1}$ for $i=1,...,N-1$ having length 2; and $\alpha
_{N}=2e_{N}$ with length 4. In this basis, the minuscule coweight is given
by $\mu _{N}=\frac{1}{2}(e_{1}+...+e_{N})$; it describes the symplectic
fundamental representation with dimension 2N. \newline
The Levi-decomposition of the Lie algebra of the symplectic gauge symmetry
reads as $so_{2}\oplus sl_{N}\oplus \boldsymbol{n}_{\pm }$ with nilpotent
algebras $\boldsymbol{n}_{\pm }=\frac{1}{2}N\left( N+1\right) _{\pm }$ which
are given by the symmetric representations of $sl_{N}$ and its conjugate.
This corresponds to cutting the first node $\alpha _{1}$ in the Dynkin
diagram of $B_{N}\simeq so_{2N+1}$ as depicted by the Figure \textbf{\ref{C6}%
}.
The adjoint action $\mathbf{\mu }$ of the minuscule coweight and the matrix
realisation of the $X_{i\bar{\imath}},X_{\left[ i\bar{j}\right] }$ and $Y^{%
\bar{\imath}i},Y^{\left[ \bar{\imath}j\right] }$ generators of the nilpotent
algebras $\boldsymbol{n}_{\pm }$ are given by%
\begin{equation}
\begin{tabular}{lll}
$X_{i\bar{\imath}}$ & $=$ & $\left \vert i\right \rangle \left \langle \bar{%
\imath}\right \vert $ \\
$Y^{\bar{\imath}i}$ & $=$ & $\left \vert \bar{\imath}\right \rangle \left
\langle i\right \vert $ \\
$X_{\left[ i\bar{j}\right] }$ & $=$ & $\left \vert i\right \rangle \left
\langle \bar{j}\right \vert -\left \vert j\right \rangle \left \langle \bar{%
\imath}\right \vert $ \\
$Y^{\left[ \bar{\imath}j\right] }$ & $=$ & $\left \vert \bar{\imath}\right
\rangle \left \langle j\right \vert -\left \vert \bar{j}\right \rangle \left
\langle i\right \vert $%
\end{tabular}%
\end{equation}%
and%
\begin{equation}
\mathbf{\mu }=\frac{1}{2}\sum_{i=1}^{N}\varrho _{i}-\frac{1}{2}\sum_{\bar{%
\imath}=\bar{1}}^{\bar{N}}\bar{\varrho}_{i}
\end{equation}%
The Lax- matrix operator $\mathcal{L}_{C}^{\mu _{N}}$ is explicitly derived
in\ section 5; it reads as follows%
\begin{equation}
\mathcal{L}_{C}^{\mu _{N}}=\left(
\begin{array}{cc}
z^{\frac{1}{2}}I_{N}+z^{-\frac{1}{2}}XY & z^{-\frac{1}{2}}X \\
z^{-\frac{1}{2}}Y & z^{-\frac{1}{2}}I_{N}%
\end{array}%
\right)  \label{Cn}
\end{equation}%
This Lax-matrix must be compared with (\ref{a2m}). The entries of X and Y
matrices in the basis $\left \{ \left \vert i\right \rangle ,\left \vert
\bar{\imath}\right \rangle \right \} $ are given by%
\begin{equation}
X=\left(
\begin{array}{ccc}
b_{1\bar{N}} & \cdots & b_{1\bar{1}} \\
\vdots & \ddots & \vdots \\
b_{N\bar{N}} & \cdots & b_{N\bar{1}}%
\end{array}%
\right) \qquad ,\qquad Y=\left(
\begin{array}{ccc}
c_{\bar{N}1} & \cdots & c_{\bar{N}N} \\
\vdots & \ddots & \vdots \\
c_{\bar{1}1} & \cdots & c_{\bar{1}N}%
\end{array}%
\right)
\end{equation}%
Multiplying (\ref{Cn}) by $z^{1/2}$, we obtain%
\begin{equation}
\mathcal{\tilde{L}}_{C}^{\mu _{N}}=\left(
\begin{array}{cc}
zI_{N}+XY & X \\
Y & I_{N}%
\end{array}%
\right)
\end{equation}

\subsubsection{D-type operators revisited}

For the case of the 4D Chern-Simons with gauge symmetry SO$_{2N+1}$; we have
three minuscule coweights $\mu _{1},\mu _{N-1}$ and $\mu _{N}$ as shown in
Table \textbf{\ref{T1}}. Therefore, we have three types of minuscule Lax
operators.%
\begin{equation}
\mathcal{L}_{D}^{\mu _{1}},\qquad \mathcal{L}_{D}^{\mu _{N-1}},\qquad
\mathcal{L}_{D}^{\mu _{N}}  \label{3d}
\end{equation}%
To get more insight into the algebraic structure underlying these
L-operators, it is interesting to recall useful relations: $\left( \mathbf{1}%
\right) $ the SO$_{2N}$ has $2N^{2}$ roots, half of them are positive and
the other half are negative. These roots have length 2 and are realised in
the weight vector basis $\left \{ e_{1},...,e_{N}\right \} $ like $\pm \left(
e_{i}\pm e_{j}\right) $ with $1\leq i<j\leq N.$ The N simple roots $\alpha
_{i}$ of SO$_{2N}$ are given by
\begin{equation}
\begin{tabular}{lllll}
$\alpha _{i}$ & $=$ & $e_{i}-e_{j}$ & for & $i=1,...,N-2$ \\
$\alpha _{N-1}$ & $=$ & $e_{N-1}-e_{N}$ &  &  \\
$\alpha _{N}$ & $=$ & $e_{N-1}+e_{N}$ &  &
\end{tabular}
\label{a3}
\end{equation}%
In this basis, the minuscule coweights, satisfying $\mu _{1}.\alpha _{1}=1$
and $\mu _{N-1}.\alpha _{N-1}=1$ as well as $\mu _{N}.\alpha _{N}=1,$ are
given by%
\begin{equation}
\begin{tabular}{lll}
$\mu _{1}$ & $=$ & $e_{1}$ \\
$\mu _{N-1}$ & $=$ & $\frac{1}{2}(e_{1}+...+e_{N-1}-e_{N})$ \\
$\mu _{N}$ & $=$ & $\frac{1}{2}(e_{1}+...+e_{N-1}+e_{N})$%
\end{tabular}
\label{m3}
\end{equation}%
The Levi-decompositions of the so$_{2N}$ Lie algebra with respect to the
three minuscule coweights are as follows; see also the Table \textbf{\ref{T1}%
},
\begin{equation}
\begin{tabular}{lll}
$\mu _{1}$ & : & $so_{2}\oplus so_{2N-2}\oplus \left( 2N-2\right) _{\pm }$
\\
$\mu _{N-1}$ & : & $so_{2}\oplus sl_{N}\oplus N\left( N-1\right) _{\pm }$ \\
$\mu _{N}$ & : & $so_{2}\oplus sl_{N}\oplus N\left( N-1\right) _{\pm }$%
\end{tabular}%
\end{equation}%
The three Lax-matrices (\ref{3d}) corresponding to these three minuscule
coweights are described below.

\  \

\begin{figure}[tbph]
\begin{center}
\includegraphics[width=6cm]{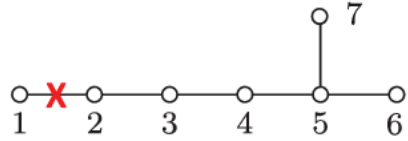}
\end{center}
\par
\vspace{-0.5cm}
\caption{Dynkin Diagram of SO$_{2N}$ illustrated for the example where $N=7$%
. By cutting the first node $\protect \alpha _{1},$ one ends with the Diagram
associated with the Dynkin diagram of the D$_{N-1}$ subalgebra and a free
node $\protect \alpha _{1}$ capturing data on the SO$_{2}$ Levi-charge and
nilpotent subalgebras.}
\label{SO71}
\end{figure}

$\mathbf{I)}$ \textbf{Vectorial coweight:} $so_{2N}\rightarrow so_{2}\oplus
so_{2N-2}\oplus \left( 2N-2\right) _{\pm }$\newline
In the decomposition of so$_{2N}$ with respect to $\mu _{1}$ corresponding
to cutting the node $\alpha _{1}$ in the Dynkin diagram of the Figure
\textbf{\ref{SO71}}, the electric charge of Wilson line is given by the
weight of the vector representation 2N which splits as $1_{+}\oplus \left(
2N-2\right) \oplus 1_{-}$.
Denoting the basis vectors like $\left \{ \left \vert +\right \rangle
,\left
\vert i\right \rangle ,\left \vert -\right \rangle \right \} $ with
label $i=1,...,2N-2,$ the Levi- constraint relations giving the generators $%
X_{i}$ and $Y^{i}$ of the nilpotent $\boldsymbol{n}_{\pm }$ are solved by%
\begin{equation}
\begin{tabular}{lll}
$X_{i}$ & $=$ & $\left \vert +\right \rangle \left \langle i\right \vert
-\left \vert i\right \rangle \left \langle -\right \vert $ \\
$Y^{i}$ & $=$ & $\left \vert i\right \rangle \left \langle +\right \vert
-\left \vert -\right \rangle \left \langle i\right \vert $%
\end{tabular}%
\end{equation}%
and%
\begin{equation}
\mathbf{\mu }=\left \vert +\right \rangle \left \langle +\right \vert
+q\sum_{i}\left \vert i\right \rangle \left \langle i\right \vert -\left
\vert -\right \rangle \left \langle -\right \vert
\end{equation}%
where $q=0$, it has been inserted for convenience. The oscillator
realisation of the Lax operator $\mathcal{L}_{D}^{vect}$ in the fundamental $%
2N$ representation is given by%
\begin{equation}
\mathcal{L}_{D}^{\mu _{1}}=\left(
\begin{array}{ccc}
z+\mathbf{b^{T}.c+}\frac{1}{4}z^{-1}\mathbf{b}^{2}\mathbf{c}^{2} & \mathbf{b}%
^{T}-\frac{1}{2}z^{-1}\mathbf{b}^{2}\mathbf{c}^{T} & \frac{1}{2}z^{-1}%
\mathbf{b}^{2} \\
\mathbf{c-}\frac{1}{2}z^{-1}\mathbf{c}^{2}\mathbf{b} & I_{2N-2}+z^{-1}%
\mathbf{b^{T}.c} & -z^{-1}\mathbf{b} \\
\frac{1}{2}z^{-1}\mathbf{c}^{2} & -z^{-1}\mathbf{c}^{T} & z^{-1}%
\end{array}%
\right)  \label{Dn1}
\end{equation}%
In this relation, we have $\mathbf{b}^{T}=\left(
0,b_{1},...,b_{2N-2},0\right) $ and $\mathbf{c}^{T}=\left(
0,c_{1},...,c_{2N-2},0\right) $ where the $(b^{i},c_{i})$ are harmonic
oscillators associated with the phase space of the vector like $D_{N}$-
system.

\  \  \  \  \

\begin{figure}[tbph]
\begin{center}
\includegraphics[width=6cm]{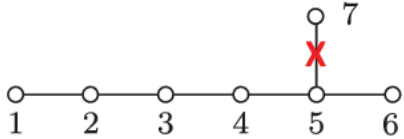}
\end{center}
\par
\vspace{-0.5cm}
\caption{Dynkin Diagram of SO$_{2N}$ illustrated for the example where N=7.
By cutting the node $\protect \alpha _{N},$ one ends with the Diagram
associated with the Levi subalgebra of A$_{N-1}$ and a free node $\protect%
\alpha _{N}$.}
\label{SO72}
\end{figure}

$\mathbf{II)}$ \textbf{Spinorial coweight}\newline
In this case, the minuscule coweight is given by $\mu _{N}$. The
Levi-decomposition is given by $so_{2}\oplus sl_{N}\oplus N\left( N-1\right)
_{\pm },$ it results from cutting the N-th node $\alpha _{N}$ in the Dynkin
diagram of D$_{N}\simeq so_{2N}$ as in the Figure \textbf{\ref{SO72}},%
\textbf{\ }
The Wilson charge representation 2N decomposes in this case like $%
N_{+1/2}\oplus N_{-1/2}.$ Using the basis vectors $\left \{ \left \vert
i\right \rangle ,\left \vert \bar{\imath}\right \rangle \right \} $ with
label $1\leq i\leq N$ and label $\bar{\imath}=2N+1-i$ taking the values $%
\bar{1}\leq \bar{\imath}\leq \bar{N},$ we solve the Levi-constraint
relations by%
\begin{equation}
\begin{tabular}{lll}
$X_{i\bar{j}}$ & $=$ & $\left \vert i\right \rangle \left \langle \bar{j}%
\right \vert -\left \vert j\right \rangle \left \langle \bar{j}\right \vert $
\\
$Y^{\bar{\imath}j}$ & $=$ & $\left \vert \bar{\imath}\right \rangle \left
\langle j\right \vert -\left \vert \bar{j}\right \rangle \left \langle
i\right \vert $%
\end{tabular}%
\end{equation}%
with%
\begin{equation}
\mathbf{\mu }=\frac{1}{2}\sum_{i}\left \vert i\right \rangle \left \langle
i\right \vert -\frac{1}{2}\sum_{\bar{\imath}}\left \vert \bar{\imath}\right
\rangle \left \langle \bar{\imath}\right \vert  \label{mm}
\end{equation}%
The calculation of the Lax matrix $\mathcal{L}_{D_{N}}^{\mu
_{N}}=e^{X}z^{\mu }e^{Y}$ with $X=b^{i\bar{j}}X_{i\bar{j}}$ and $Y=c_{i\bar{j%
}}Y^{\bar{\imath}j}$ leads to%
\begin{equation}
\mathcal{L}_{D_{N}}^{\mu _{N}}=\left(
\begin{array}{cc}
z^{\frac{1}{2}}+z^{-\frac{1}{2}}BC & z^{-\frac{1}{2}}B \\
z^{-\frac{1}{2}}C & z^{-\frac{1}{2}}%
\end{array}%
\right)  \label{Dn2}
\end{equation}%
In this relation, the N$\times $\={N} matrix $B$ and the \={N}$\times $N
matrix $C$ are given by:
\begin{equation}
B=\left \{
\begin{array}{ccc}
b^{i\bar{j}} & \qquad & i<j \\
-b^{j\bar{\imath}} & \qquad & j<i \\
0 & \qquad & i=j%
\end{array}%
\right. ,\qquad C=\left \{
\begin{array}{ccc}
c_{\bar{j}i} & \qquad & i<j \\
-c_{\bar{\imath}j} & \qquad & j<i \\
0 & \qquad & i=j%
\end{array}%
\right.
\end{equation}%
In matrix notations, we have$\ $%
\begin{equation}
B=\left(
\begin{array}{ccccc}
b_{1\bar{N}} & b_{1\bar{N}-\bar{1}} & \cdots & b_{1\bar{2}} & 0 \\
b_{2\bar{N}} & b_{2\bar{N}-\bar{1}} & \cdots & 0 & -b_{1\bar{2}} \\
\vdots & \vdots & \ddots & \vdots & \vdots \\
b_{N-1,\bar{N}} & 0 & \cdots & -b_{2,\bar{N}-\bar{1}} & -b_{1,\bar{N}-\bar{1}%
} \\
0 & -b_{N-1,\bar{N}} & \cdots & -b_{2\bar{N}} & -b_{1\bar{N}}%
\end{array}%
\right)
\end{equation}%
and%
\begin{equation}
C=\left(
\begin{array}{ccccc}
c_{\bar{N}1} & c_{\bar{N}2} & \cdots & c_{\bar{N},N-1} & 0 \\
c_{\bar{N}-\bar{1},1} & c_{\bar{N}-\bar{1},2} & \cdots & 0 & -c_{\bar{N}-%
\bar{1},N} \\
\vdots & \vdots & \ddots & \vdots & \vdots \\
c_{\bar{2}1} & 0 & \cdots & -c_{\bar{2},\bar{N}-\bar{1}} & -c_{\bar{2},N} \\
0 & -c_{\bar{1}2} & \cdots & -c_{\bar{1}N-1} & c_{\bar{1}N}%
\end{array}%
\right)
\end{equation}

\  \  \  \  \

\begin{figure}[tbph]
\begin{center}
\includegraphics[width=6cm]{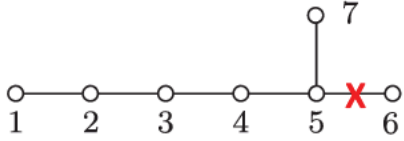}
\end{center}
\par
\vspace{-0.5cm}
\caption{Dynkin Diagram of SO$_{2N}$ illustrated for the example where N=7.
By cutting the node $\protect \alpha _{N-1},$ one ends with the Diagram
associated with the Levi subalgebra of A'$_{N-1}$ and a free node $\protect%
\alpha _{N-1}$.}
\label{SO73}
\end{figure}

$\mathbf{III)}$ \textbf{Cospinorial coweight}\newline
In this case, the minuscule coweight is given by $\mu _{N-1}$. The
Levi-decomposition is given by $so_{2}\oplus sl_{N}^{\prime }\oplus N\left(
N-1\right) _{\pm };$\ it corresponds to cutting the node $\alpha _{N-1}$ in
the Dynkin diagram of the Figure \textbf{\ref{SO73}. }
Here, the $A_{N-1}^{\prime }\simeq sl_{N}^{\prime }$ is isomorphic to the $%
A_{N-1}\simeq sl_{N}$ appearing in the decomposition using the coweight $\mu
_{N}$ considered above. In the present case, the $sl_{N}^{\prime }$ results
from cutting the N-th node $\alpha _{N-1}$ in the Dynkin diagram of $so_{2N}$
while cutting the node $\alpha _{N}$ yields $sl_{N}$. As such, the simple
root systems of the $sl_{N}^{\prime }$ and $sl_{N}$ are isometric and are
given by
\begin{equation}
\begin{tabular}{lll}
$sl_{N}^{\prime }$ & : & $\left \{ \alpha _{1},...,\alpha _{N-2},\alpha
_{N}\right \} $ \\
$sl_{N}$ & : & $\left \{ \alpha _{1},...,\alpha _{N-2},\alpha _{N-1}\right
\} $%
\end{tabular}%
\end{equation}%
These two systems are related to each other by the outer-automorphism
symmetry $\mathbb{Z}_{2}$ acting by permutation of the simple roots $\alpha
_{N-1}$ and $\alpha _{N}$ as follows%
\begin{equation}
\mathbb{Z}_{2}:\left \{
\begin{array}{lllll}
\alpha _{i} & \rightarrow & \alpha _{i} & \text{ \ for \ } & 1\leq i\leq N-2
\\
\alpha _{N-1} & \rightarrow & \alpha _{N} &  &  \\
\alpha _{N} & \rightarrow & \alpha _{N-1} &  &
\end{array}%
\right.
\end{equation}%
Notice that the above $\mathbb{Z}_{2}$ discrete symmetry acts non trivially
on the 2N quantum states $\left \vert a\right \rangle ,$ with label $1\leq
a\leq 2N,$ propagating along the Wilson line. As this electric line carries
a vector-like charge given by the SO$_{2N}$ representation; and by using the
decomposition $2N=N_{+1/2}\oplus N_{-1/2}$ respectively labeled by $%
\left
\vert i\right \rangle $ and $\left \vert \bar{\imath}\right \rangle $
with $1\leq i\leq N$ and $\bar{\imath}=2N+1-i$ taking the values $\bar{1}%
\leq \bar{\imath}\leq \bar{N},$ we end up with the following transformations%
\begin{equation}
\mathbb{Z}_{2}:\left \{
\begin{array}{lllll}
\left \vert i\right \rangle & \rightarrow & \left \vert i\right \rangle &
\text{ \ for \ } & 1\leq i\leq N-2 \\
\left \vert N-1\right \rangle & \rightarrow & \left \vert N\right \rangle &
&  \\
\left \vert N\right \rangle & \rightarrow & \left \vert N-1\right \rangle &
&
\end{array}%
\right.  \label{ch}
\end{equation}%
Under this transformation, the charge operator (\ref{mm}) is preserved
because the sums $\sum \left \vert i\right \rangle \left \langle
i\right
\vert $ and $\sum \left \vert \bar{\imath}\right \rangle
\left
\langle \bar{\imath}\right \vert $ are not affected by (\ref{ch}).
The same invariance holds for the linear expansions $X=b^{i\bar{j}}X_{i\bar{j%
}}$ and $Y=c_{i\bar{j}}Y^{\bar{\imath}j}$. So, the Lax-matrix $\mathcal{L}%
_{D}^{\mu _{N-1}}$ is the same as $\mathcal{L}_{D}^{\mu _{N}}$ namely%
\begin{equation}
\mathcal{L}_{D}^{\mu _{N-1}}=\left(
\begin{array}{cc}
z^{\frac{1}{2}}+z^{-\frac{1}{2}}BC & z^{-\frac{1}{2}}B \\
z^{-\frac{1}{2}}C & z^{-\frac{1}{2}}%
\end{array}%
\right)  \label{Dn3}
\end{equation}

\subsubsection{Exceptional Lax operators}

As there is no minuscule coweight in the exceptional Lie algebra E$_{8}$, we
have minuscule Lax-operators only for the 4D exceptional Chern-Simons with
gauge symmetries E$_{6}$ and E$_{7}$. From the classification Table \textbf{%
\ref{T1}}, we learn that $\left( \mathbf{i}\right) $ the E$_{6}$ gauge model
has two minuscule L-operators $\mathcal{L}_{E_{6}}^{\mu _{1}}$ and $\mathcal{%
L}_{E_{6}}^{\mu _{5}}.$ $\left( \mathbf{ii}\right) $ the E$_{7}$ gauge model
has one minuscule $\mathcal{L}_{E_{7}}^{\mu _{1}}$. They are described here
below.

\  \  \

\begin{figure}[tbph]
\begin{center}
\includegraphics[width=8cm]{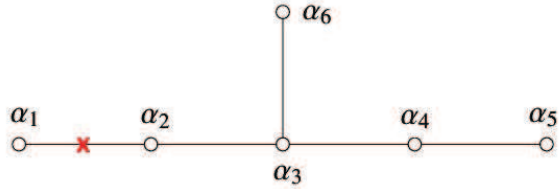}
\end{center}
\par
\vspace{-0.5cm}
\caption{Dynkin Diagram of E$_{6}$ having six nodes labeled by the six
simple roots $\protect \alpha _{i}$. The cross $\left( \times \right) $
indicates the roots used in the Levi decomposition with Levi subalgebra as $%
so_{10}\oplus so_{2}.$}
\label{E6}
\end{figure}

$\mathbf{I)}$ \textbf{Lax-matrices} $\mathcal{L}_{E_{6}}^{\mu _{1}}$ \textbf{%
and} $\mathcal{L}_{E_{6}}^{\mu _{5}}$: \newline
The Lax operator of E$_{6}$-type for the coweight $\mu _{1}$ is associated
with the Levi-decomposition $E_{6}\rightarrow so_{10}\oplus so_{2}\oplus
16_{+}\oplus 16_{-}$. In the diagrammatic language, this corresponds to
omitting the first node in the Dynkin diagram of $E_{6}$ given by the Figure
\textbf{\ref{E6}}.
This node, labeled by simple root $\alpha _{1},$ corresponds to the
fundamental $\mathbf{27}$\ representation of E$_{6}$ with quantum states $%
\left \vert \varphi \right \rangle $ propagating along the Wilson line.
These 27 quantum states decompose with respect to the Levi-subalgebra as
follows
\begin{equation}
27=1_{-4/3}\oplus 10_{+2/3}\oplus 16_{-1/3}
\end{equation}%
So, we split the 27 states of the\ representation $\mathbf{27}$ like: $%
\left( i\right) $ a state $\left \vert 0\right \rangle $ denoting the
singlet $1_{-4/3}$; $\left( ii\right) $ ten states $\left \{ \left \vert
i\right \rangle \right \} _{1\leq i\leq 10}$ designating the ten-uplet $%
10_{+2/3}$ and $\left( iii\right) $ sixteen states $\left \{ \left \vert
\beta \right \rangle \right \} _{1\leq \beta \leq 16}$ representing the $%
16_{-1/3}$. Using these states, we solve the Levi-constraint relations for
generators of the nilpotent subalgebras $\mathbf{16}_{\pm }$ as
\begin{equation}
\begin{tabular}{lll}
$X_{\beta }$ & $=$ & $\left \vert \beta \right \rangle \left \langle 0\right
\vert +\left( \Gamma ^{i}\right) _{\beta \gamma }\left \vert i\right \rangle
\left \langle \gamma \right \vert $ \\
$Y^{\beta }$ & $=$ & $\left \vert 0\right \rangle \left \langle \beta \right
\vert +\left( \Gamma ^{i}\right) _{\beta \gamma }\left \vert \gamma \right
\rangle \left \langle i\right \vert $%
\end{tabular}
\label{e6xy}
\end{equation}%
and%
\begin{equation}
\mathbf{\mu }=-\frac{4}{3}\left \vert 0\right \rangle \left \langle 0\right
\vert +\frac{2}{3}\sum_{i}\left \vert i\right \rangle \left \langle i\right
\vert -\frac{1}{3}\sum_{\beta }\left \vert \beta \right \rangle \left
\langle \beta \right \vert  \label{e6m}
\end{equation}%
where $\Gamma ^{i}$ are $so_{10}$ Dirac-like matrices satisfying the
euclidian Clifford algebra. Putting these relations back into $\mathcal{L}%
_{E_{6}}^{\mu _{1}}=e^{X}z^{\mu }e^{Y}$, we obtain after tedious but
straightforward algebra the following expression,%
\begin{equation}
\mathcal{L}_{E_{6}}^{\mu _{1}}=\left(
\begin{array}{ccc}
z^{-\frac{4}{3}} & z^{-\frac{4}{3}}W_{j} & z^{-\frac{4}{3}}c_{\beta } \\
z^{-\frac{4}{3}}V^{i} & z^{\frac{2}{3}}\delta _{j}^{i}+z^{-\frac{4}{3}%
}V^{i}W_{j}+z^{-\frac{1}{3}}B_{\alpha }^{i}C_{j}^{\alpha } & z^{-\frac{4}{3}%
}V^{i}c_{\beta }+z^{-\frac{1}{3}}B_{\beta }^{i} \\
z^{-\frac{4}{3}}b^{\alpha } & z^{-\frac{4}{3}}b^{\alpha }W_{j}+z^{-\frac{1}{3%
}}C_{j}^{\alpha } & z^{-\frac{1}{3}}\delta _{\beta }^{\alpha }+z^{-\frac{4}{3%
}}b^{\alpha }c_{\beta }%
\end{array}%
\right)  \label{E61}
\end{equation}%
where we have defined the following quantities in terms of the oscillator
degrees of freedom $b^{\alpha }$ and $c_{\beta }.$%
\begin{equation}
\begin{tabular}{lll}
$B_{\beta }^{i}=b^{\gamma }\Gamma _{\gamma \beta }^{i}$ & $,$ & $V^{i}=\frac{%
1}{2}b^{\alpha }\left( \Gamma ^{i}\right) _{\alpha \beta }b^{\beta }$ \\
$C_{j}^{\alpha }=c_{\gamma }\Gamma _{j}^{\gamma \alpha }$ & $,$ & $W_{i}=%
\frac{1}{2}c_{\alpha }\left( \Gamma _{i}\right) ^{\alpha \beta }c_{\beta }$%
\end{tabular}%
\end{equation}%
The Lax operator $\mathcal{L}_{E_{6}}^{\mu _{5}}$ associated to the coweight
$\mu _{5}$ is obtained in a similar way as before; but instead of cutting
the first node $\alpha _{1}$ in the Dynkin diagram of E$_{6}$, we need to
cut the fifth $\alpha _{5}$. Because this node corresponds to the
anti-fundamental representation $\overline{\mathbf{27}},$ it is realised in
the same basis as in (\ref{e6xy}-\ref{e6m}) but with $\mathbf{\mu }$
replaced with the opposite $-\mathbf{\mu }$. This feature leads to the
relationship%
\begin{equation}
\mathcal{L}_{E_{6}}^{\mu _{5}}\left( z\right) =\mathcal{L}_{E_{6}}^{\mu
_{1}}\left( 1/z\right)  \label{E62}
\end{equation}%
The properties of the\ exceptional Lax operators $\mathcal{L}_{E_{6}}^{\mu }$
were briefly outlined in \cite{4D} and its detailed derivation can be found
in \cite{1G}.

\  \  \

$\mathbf{II)}$ \textbf{Lax-matrix }$\mathcal{L}_{E_{7}}$\newline
The gauge symmetry group $E_{7}$ of the 4D Chern-Simons theory is
characterized by one minuscule coweight $\mu _{1}$ corresponding to the
fundamental representation $\mathbf{56}$. As shown on the Tables \textbf{\ref%
{T1}}-\textbf{\ref{T2}}, the Levi-subalgebra of $E_{7}$ contains its
subalgebra E$_{6}$ and the nilpotent $\boldsymbol{n}_{\pm }$ are given by $%
\mathbf{27}_{\pm }$. This corresponds to cutting the node $\alpha _{6}$ as
depicted by the Figure \textbf{\ref{E7}.}
\begin{figure}[tbph]
\begin{center}
\includegraphics[width=8cm]{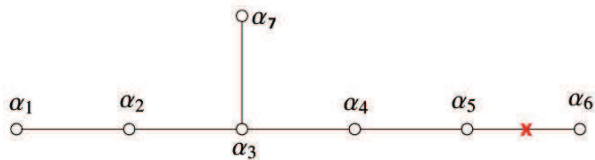}
\end{center}
\par
\vspace{-0.5cm}
\caption{Dynkin Diagram of E$_{7}$ having seven nodes labeled by the simple
roots $\protect \alpha _{i}$. The cross $\left( \times \right) $ indicates
the roots used in the Levi decomposition with Levi subgroup $SO(2)\times
E_{6}.$}
\end{figure}
The quantum states $\left \vert a\right \rangle $ running on the
electrically charged Wilson line are described by the representation $%
\mathbf{56}$ that decomposes as follows
\begin{equation}
\mathbf{56=1}_{3/2}\oplus \mathbf{27}_{1/2}\oplus \overline{27}_{-1/2}\oplus
\mathbf{\bar{1}}_{-3/2}
\end{equation}%
As such, the 56 states $\left \vert a\right \rangle $ can be splitted as $%
\left \vert 0_{+}\right \rangle ,\left \vert \beta _{+}\right \rangle
,\left
\vert \beta _{-}\right \rangle ,\left \vert 0_{-}\right \rangle $
where the labels take the values $\beta _{+}=1,...,27$ and $\beta _{-}=\bar{1%
},...,\overline{27}.$ The generators of $\boldsymbol{n}_{\pm }$ solving the
Levi-conditions are given by%
\begin{equation}
\begin{tabular}{lll}
$X_{\beta }$ & $=$ & $\left \vert 0_{+}\right \rangle \left \langle \beta
_{+}\right \vert +\left \vert \delta _{+}\right \rangle \Gamma _{\beta
}^{\delta _{+}\gamma _{-}}\left \langle \gamma _{-}\right \vert +\left \vert
\beta _{-}\right \rangle \left \langle 0_{-}\right \vert $ \\
$Y^{\beta }$ & $=$ & $\left \vert 0_{-}\right \rangle \left \langle \beta
_{-}\right \vert +\left \vert \gamma _{-}\right \rangle \bar{\Gamma}_{\gamma
_{-}\delta _{+}}^{\beta }\left \langle \delta _{+}\right \vert +\left \vert
\beta _{+}\right \rangle \left \langle 0_{+}\right \vert $%
\end{tabular}%
\end{equation}%
such that the $\Gamma _{\beta }^{\delta _{+}\gamma _{-}}$ and $\bar{\Gamma}%
_{\gamma _{-}\delta _{+}}^{\beta }$ are tri-coupling objects of the E$_{6}$
representation theory \cite{1G}. The Levi- charge operator is given by%
\begin{equation}
\mathbf{\mu }=\frac{3}{2}\left \vert 0_{+}\right \rangle \left \langle
0_{+}\right \vert +\frac{1}{2}\sum_{\beta _{+}}\left \vert \beta _{+}\right
\rangle \left \langle \beta _{+}\right \vert -\frac{1}{2}\sum_{\beta
_{-}}\left \vert \beta _{-}\right \rangle \left \langle \beta _{-}\right
\vert -\frac{3}{2}\left \vert 0_{-}\right \rangle \left \langle 0_{-}\right
\vert
\end{equation}%
The derivation of the Lax-matrix $\mathcal{L}_{E_{7}}^{\mu _{1}}$ is a
little bit technical and \textrm{cumbersome}, we choose to represent it in
the basis $\left \vert 0_{+}\right \rangle ,\left \vert \beta
_{+}\right
\rangle ,\left \vert \beta _{-}\right \rangle ,\left \vert
0_{-}\right
\rangle $ as follows
\begin{equation}
\mathcal{L}_{E_{7}}^{\mu _{1}}=\left(
\begin{array}{cccc}
L_{0_{+}}^{0_{+}} & L_{0_{+}}^{\beta _{+}} & L_{0_{+}}^{\beta _{-}} &
L_{0_{+}}^{0-} \\
L_{\alpha _{+}}^{0_{+}} & L_{\alpha _{+}}^{\beta _{+}} & L_{\alpha
_{+}}^{\beta _{-}} & L_{\alpha _{+}}^{0-} \\
L_{\alpha _{-}}^{0_{+}} & L_{\alpha _{-}}^{\beta _{+}} & L_{\alpha
_{-}}^{\beta _{-}} & L_{\alpha _{-}}^{0-} \\
L_{0-}^{0_{+}} & L_{0-}^{\beta _{+}} & L_{0-}^{\beta _{-}} & L_{0-}^{0-}%
\end{array}%
\right)  \label{E7}
\end{equation}%
The diagonal entries of this Lax-matrix are given by%
\begin{equation}
\begin{tabular}{lll}
$L_{0_{+}}^{0_{+}}$ & $=$ & $z^{\frac{3}{2}}+z^{\frac{1}{2}}b^{\alpha
_{+}}c_{\alpha _{+}}+z^{-\frac{1}{2}}S^{\alpha _{-}}R_{\alpha _{-}}+z^{-%
\frac{3}{2}}\mathcal{EF}$ \\
$L_{\alpha _{+}}^{\beta _{+}}$ & $=$ & $z^{\frac{1}{2}}\delta _{\alpha
_{+}}^{\beta _{+}}+z^{-\frac{1}{2}}T_{\alpha _{+}}^{\gamma _{-}}J_{\gamma
_{-}}^{\beta _{+}}+z^{-\frac{3}{2}}S_{\alpha _{+}}R^{\beta _{+}}$ \\
$L_{\alpha _{-}}^{\beta _{-}}$ & $=$ & $z^{-\frac{1}{2}}\delta _{\alpha
_{-}}^{\beta _{-}}+z^{-\frac{3}{2}}b_{\alpha _{-}}c^{\beta _{-}}$ \\
$L_{0-}^{0-}$ & $=$ & $z^{-\frac{3}{2}}$%
\end{tabular}%
\end{equation}%
with $\mathcal{E}=b^{\alpha }b^{\beta }b^{\beta }\Gamma _{\alpha \beta
\gamma }$ and $\mathcal{F}=\bar{\Gamma}^{\alpha \beta \gamma }c_{\alpha
}c_{\beta }c_{\gamma }$. The other Lax-matrix entries are as listed below%
\begin{equation}
\begin{tabular}{lll}
${\small L}_{{\small 0}_{+}}^{{\small \beta }_{+}}$ & $=$ & ${\small z}^{%
\frac{1}{2}}{\small b}^{\beta _{+}}{\small +z}^{-\frac{1}{2}}{\small S}^{%
{\small \alpha }_{-}}{\small J}_{{\small \alpha }_{-}}^{{\small \beta }_{+}}%
{\small +z}^{-\frac{3}{2}}\mathcal{E}R^{{\small \beta }_{+}}$ \\
${\small L}_{{\small 0}_{+}}^{{\small \beta }_{-}}$ & $=$ & $z^{-\frac{1}{2}%
}S^{\beta _{-}}+z^{-\frac{3}{2}}\mathcal{E}c^{\beta _{-}}$ \\
${\small L}_{{\small \alpha }_{+}}^{{\small 0}_{+}}$ & $=$ & $z^{\frac{1}{2}%
}c_{{\small \alpha }_{+}}+z^{-\frac{1}{2}}T_{{\small \alpha }_{+}}^{{\small %
\gamma }_{-}}R_{{\small \gamma }_{-}}+z^{-\frac{3}{2}}S_{{\small \alpha }%
_{+}}\mathcal{F}$%
\end{tabular}%
\end{equation}%
and%
\begin{equation}
\begin{tabular}{lll}
$L_{\alpha _{-}}^{0_{+}}$ & $=$ & $z^{-\frac{1}{2}}R_{\alpha _{-}}+z^{-\frac{%
3}{2}}b_{\alpha _{-}}\mathcal{F}$ \\
$L_{\alpha _{+}}^{\beta _{-}}$ & $=$ & $z^{-\frac{1}{2}}T_{\alpha
_{+}}^{\beta _{-}}+z^{-\frac{3}{2}}S_{\alpha _{+}}c^{\beta _{-}}$ \\
$L_{\alpha _{-}}^{\beta _{+}}$ & $=$ & $z^{-\frac{1}{2}}J_{\alpha
_{-}}^{\beta _{+}}+z^{-\frac{3}{2}}b_{\alpha _{-}}R^{\beta _{+}}$%
\end{tabular}%
\end{equation}%
as well as%
\begin{equation}
\begin{tabular}{lllllll}
$L_{0_{+}}^{0-}$ & $=$ & $z^{-\frac{3}{2}}\mathcal{E}$ & , & $L_{0-}^{0_{+}}$
& $=$ & $z^{-\frac{3}{2}}\mathcal{F}$ \\
$L_{\alpha _{+}}^{0-}$ & $=$ & $z^{-\frac{3}{2}}S_{\alpha _{+}}$ & , & $%
L_{0-}^{\beta _{+}}$ & $=$ & $z^{-\frac{3}{2}}R^{\beta _{+}}$ \\
$L_{\alpha _{-}}^{0-}$ & $=$ & $z^{-\frac{3}{2}}b_{\alpha _{-}}$ & , & $%
L_{0-}^{\beta _{-}}$ & $=$ & $z^{-\frac{3}{2}}c^{\beta _{-}}$%
\end{tabular}%
\end{equation}%
More details concerning the derivation of this operator from the 4D Chern-
Simons theory can be found in \cite{1G}.

\section{B-type Lax operators}

In this section, we calculate the minuscule Lax operator $\mathcal{L}_{B}(z)$%
\ from the 4D Chern- Simons theory with gauge symmetry $SO_{2N+1}.$ The $%
\mathcal{L}_{B}^{\mu }$ is determined by using the formula $e^{X}z^{\mu
}e^{Y}$. In this relation, the $\mu $ is the minuscule coweight of the
underlying Lie algebra $B_{N}$ of the gauge symmetry. The $X=b^{i}X_{i}$ and
$Y=c_{i}X^{i}$ are $\left( 2N+1\right) \times \left( 2N+1\right) $ matrices
valued in the nilpotent algebras $\boldsymbol{n}_{-}$ and $\boldsymbol{n}%
_{+} $ appearing in the Levi-decomposition of $B_{N}\sim so_{2N+1}$ with
respect to $\mu ,$ namely
\begin{equation}
\begin{tabular}{lll}
$so_{2N+1}$ & $=$ & $\boldsymbol{l}_{\mu }\oplus \boldsymbol{n}_{+}\oplus
\boldsymbol{n}_{-}$ \\
$\  \  \  \boldsymbol{l}_{\mu }$ & $=$ & $so_{2}\oplus so_{2N-1}$ \\
\  \  \ $\boldsymbol{n}_{\pm }$ & $=$ & $\left( \boldsymbol{2N-1}\right) _{\pm
1}$%
\end{tabular}
\label{b1}
\end{equation}%
Dimensions and ranks of the algebras appearing in (\ref{b1}) can be directly
read from the following decompositions%
\begin{equation}
\begin{tabular}{lllll}
$\dim $ & : & $N\left( 2N+1\right) $ & $=$ & $1+\left( 2N-1\right) \left(
N-1\right) +\left( 2N-1\right) _{\pm 1}$ \\
$rank$ & : & $\  \  \  \  \ N$ & $=$ & $1+\left( N-1\right) $%
\end{tabular}%
\end{equation}%
Recall that the finite dimensional Lie algebra $B_{N}$ has one minuscule
coweight $\mu _{1}=e_{1}$ (denoted here as $\mu $); it is the dual of the
simple root $\alpha _{1}=e_{1}-e_{2}$ ($\mu _{1}.\alpha _{i}=\delta _{1i}$);
and corresponds to the first node of the Dynkin diagram of $B_{N}$. For an
illustration, see the Figure \textbf{\ref{B6}}.\newline
The nilpotent sub-algebras are generated by X$_{i}$ and Y$^{i}$ obeying the
following commutation relations
\begin{equation}
\begin{tabular}{lll}
$\left[ \mathbf{\mu },X_{i}\right] $ & $=$ & $+X_{i}$ \\
$\left[ \mathbf{\mu },Y^{i}\right] $ & $=$ & $-Y^{i}$ \\
$\left[ X_{i},Y^{i}\right] $ & $=$ & $\delta _{i}^{i}\mu $%
\end{tabular}
\label{lce}
\end{equation}%
where $\mathbf{\mu }$\ stands for the adjoint action of the minuscule
coweight. The explicit realisation of this algebra in terms of the harmonic
oscillators of the phase space of the L-operators\ is investigated below.

\subsection{Solving Levi-constraint relations}

To solve (\ref{lce}), we need to define the Levi-decomposition of the
vectorial representation $\mathbf{2N+1}$ of the Lie algebra $so_{2N+1}$ and
related objects. Vector states of \underline{$\mathbf{2N+1}$} propagate on
the Wilson line interacting with the 't Hooft line with magnetic charge $\mu
$. Under the Levi-decomposition, the real $\mathbf{2N+1}$ representation of
so$_{2N+1}$ splits as direct sum of representations of $\boldsymbol{l}_{\mu
}=so_{2}\oplus so_{2N-1}$. In fact, we have $\mathbf{2N+1}=\mathbf{%
2_{0}\oplus (2N-1)_{0}}$ where the zero label refers to the charge under the
group $SO\left( 2\right) \mathbf{.}$ By using the isomorphism $SO\left(
2\right) \sim U(1),$ we can put this decomposition to the following form
\begin{subequations}
\begin{equation}
\mathbf{2N+1}=\mathbf{1}_{+1}\oplus \left( \mathbf{2N-1}\right) _{0}\oplus
\mathbf{1}_{-1}  \label{B1}
\end{equation}%
where we have substituted with $\mathbf{2_{0}}=\mathbf{1}_{+1}\oplus \mathbf{%
1}_{-1}$. For convenience, we use the kets $\left( \left \vert
+\right
\rangle ,\left \vert i\right \rangle ,\left \vert -\right \rangle
\right) $ with $i=1,...,2N-1$ and the bras $\left( \left \langle
+\right
\vert ,\left
\langle i\right \vert ,\left \langle -\right \vert
\right) $ to denote the vector basis of the fundamental representation $(%
\mathbf{2N+1)} $ and its dual. We have
\end{subequations}
\begin{equation}
\left \vert \mathbf{2N+1}\right \rangle =\left(
\begin{array}{c}
\left \vert +\right \rangle \\
\left \vert \mathbf{2N-1}\right \rangle \\
\left \vert -\right \rangle%
\end{array}%
\right)  \label{B3}
\end{equation}%
where $\left \vert \pm \right \rangle $ refer to the two complex singlets $%
\mathbf{1}_{\pm 1}$ in the decomposition (\ref{B1}) and $\left \vert \mathbf{%
2N-1}\right \rangle $ represents the $2N-1$ states $\left \{ \left \vert
i\right \rangle \right \} _{1\leq i\leq 2N-1}.$ This basis is characterized
by the following orthogonality relations $\left \langle +|-\right \rangle
=\left \langle +|i\right \rangle =\left \langle i|-\right \rangle =0$ and $%
\left \langle +|+\right \rangle =\left \langle -|-\right \rangle =1,$ as
well as $\left \langle i|j\right \rangle =\delta _{ij}.$\newline
The next step is to find the adjoint action $\mathbf{\mu }$ of the minuscule
coweight on the fundamental representation. This is a hermitian charge
operator acting on the quantum states generated by (\ref{B3}). It can be
represented like
\begin{equation}
\mathbf{\mu }=\varrho _{+}+q\Pi -\varrho _{\mathbf{-}}
\end{equation}%
where $\varrho _{\pm }$ are the projectors on the representation sub-spaces $%
\mathbf{1}_{\pm 1}$ appearing in (\ref{B1}) and $\Pi $ is the projector on $%
\left( \mathbf{2N-1}\right) _{0}$. These projectors read in terms of the
bras/kets as follows
\begin{equation}
\begin{tabular}{lll}
$\varrho _{+}$ & $=$ & $\left \vert +\right \rangle \left \langle +\right
\vert $ \\
$\varrho _{-}$ & $=$ & $\left \vert -\right \rangle \left \langle -\right
\vert $ \\
$\varrho _{i}$ & $=$ & $\left \vert i\right \rangle \left \langle i\right
\vert $%
\end{tabular}%
\end{equation}%
and
\begin{equation}
\Pi =\sum_{i=1}^{2N-1}\varrho _{i}
\end{equation}%
Notice here that the charge $q$ vanishes ($q=0)$ because the $\left \vert
\mathbf{2N-1}\right \rangle $ is chargeless. Now, we move to working out the
explicit expressions of the generators $X_{i}$ and $Y^{i}$ of the nilpotent
algebras $\boldsymbol{n}_{+}$ and $\boldsymbol{n}_{-}$ in the basis \ref{B3}%
. They are obtained by solving the Levi-constraint relations (\ref{lce}). By
taking $X_{i}$ and $Y^{i}$ like%
\begin{equation}
\begin{tabular}{lll}
$X_{i}$ & $=$ & $x_{1}\left \vert +\right \rangle \left \langle i\right
\vert +x_{2}\left \vert i\right \rangle \left \langle -\right \vert $ \\
$Y^{i}$ & $=$ & $y_{1}\left \vert i\right \rangle \left \langle +\right
\vert +y_{2}\left \vert -\right \rangle \left \langle i\right \vert $%
\end{tabular}
\label{xry}
\end{equation}%
with $x_{1,2}$ and $y_{1,2}$ non vanishing arbitrary numbers, then using $%
\mathbf{\mu }=\varrho _{+}-\varrho _{\mathbf{-}},$ we have $\left[ \mathbf{%
\mu },X_{i}\right] =+X_{i}$ and $\left[ \mathbf{\mu },Y^{i}\right] =-Y^{i}$
as well as $\left[ X_{i},Y^{i}\right] =\delta _{i}^{i}\mathbf{\mu }$
provided the following conditions are satisfied: $\left( \mathbf{i}\right) $
$x_{2}y_{2}=x_{1}y_{1}$ and $\left( \mathbf{ii}\right) $ $%
x_{1}y_{1}=x_{2}y_{2}=1.$ These conditions can be solved by taking
\begin{equation}
x_{1}y_{1}=1\qquad ,\qquad x_{2}=y_{2}=\pm 1
\end{equation}%
Below, we take $x_{2}=y_{2}=-1$. With these generators, we can express the $%
X $ and $Y$ matrices appearing in the Lax operator that we want to
calculate; we have
\begin{equation}
X=b^{i}X_{i}\in n_{+}\qquad ,\qquad Y=c_{i}Y^{i}\in n_{-}
\end{equation}%
In these expansions, the $b^{i}$ and the $c_{i}$ variables are the phase
space coordinates of the $\mathcal{L}_{B}$-operator; they are treated here
classically but they can be promoted to operators without ambiguity. This is
because in the formula $e^{X}z^{\mu }e^{Y},$ the $b^{i}$'s are in the left
and the $c_{i}$'s are in the right in agreement with the Wick theorem.

\subsection{Building the operator $\mathcal{L}_{B}$}

We begin by calculating the exponentials $e^{X}$ and $e^{Y}$ by using the
expansion $e^{A}=\sum A^{n}/n!.$ Based on the eqs (\ref{xry}), we compute
the powers of the $X_{i}$ and $Y^{i}$ generators. We find after some
calculations that%
\begin{equation}
\begin{tabular}{lll}
$X_{i}X_{j}$ & $=$ & $\left \vert +\right \rangle \delta _{ij}\left \langle
-\right \vert $ \\
$Y^{i}Y^{j}$ & $=$ & $\left \vert -\right \rangle \delta ^{ij}\left \langle
+\right \vert $%
\end{tabular}%
\end{equation}%
and
\begin{equation}
\begin{tabular}{lll}
$X_{i}X_{j}X_{k}$ & $=$ & $0$ \\
$Y^{i}Y^{j}Y^{k}$ & $=$ & $0$%
\end{tabular}%
\end{equation}%
From these relations and the expansions $X=b^{i}X_{i}$ and $Y=c_{i}Y^{i}$,
we deduce that $X%
{{}^2}%
=\mathbf{b}^{2}\left \vert +\right \rangle \left \langle -\right \vert $
with $\mathbf{b}^{2}=b^{i}\delta _{ij}b^{j}$ and $Y%
{{}^2}%
=\mathbf{c}^{2}\left \vert -\right \rangle \left \langle +\right \vert $
with $\mathbf{c}^{2}=c_{i}\delta ^{ij}c_{j}$. We also have $%
e^{X}=1+X+X^{2}/2 $ and $e^{Y}=1+Y+Y^{2}/2$ because $X^{3}=Y^{3}=0.$ These
features lead to the following polynomial-like expansion
\begin{equation}
\mathcal{L}_{B}=\left( 1+X+\frac{1}{2}X^{2}\right) z^{\mathbf{\mu }}\left(
1+Y+\frac{1}{2}Y^{2}\right)
\end{equation}%
reading explicitly as
\begin{equation}
\begin{tabular}{lll}
$\mathcal{L}_{B}$ & $=$ & $z^{\mathbf{\mu }}+Xz^{\mathbf{\mu }}+z^{\mathbf{%
\mu }}Y+$ \\
&  & $Xz^{\mathbf{\mu }}Y+\frac{1}{2}X^{2}z^{\mathbf{\mu }}+\frac{1}{2}z^{%
\mathbf{\mu }}Y^{2}+$ \\
&  & $\frac{1}{2}X^{2}z^{\mathbf{\mu }}Y+\frac{1}{2}Xz^{\mathbf{\mu }}Y^{2}+%
\frac{1}{4}X^{2}z^{\mathbf{\mu }}Y^{2}$%
\end{tabular}%
\end{equation}%
Replacing $z^{\mathbf{\mu }}$ with
\begin{equation}
z^{\mathbf{\mu }}=z\varrho _{+}+\Pi +z^{-1}\varrho _{-}
\end{equation}%
and taking advantage of the properties $X_{i}\varrho _{+}=0$ and $\varrho
_{+}Y^{i}=0$ as well as%
\begin{equation}
\begin{tabular}{llllll}
$X\varrho _{+}$ & $=0$ & $\qquad ,\qquad $ & $\Pi X%
{{}^2}%
$ & $=X%
{{}^2}%
\Pi $ & $=0$ \\
$\varrho _{+}Y$ & $=0$ & $\qquad ,\qquad $ & $\Pi Y%
{{}^2}%
$ & $=Y%
{{}^2}%
\Pi $ & $=0$ \\
$X%
{{}^2}%
\varrho _{+}$ & $=0$ & $\qquad ,\qquad $ & $\varrho _{+}Y%
{{}^2}%
$ & $=0$ &
\end{tabular}%
\end{equation}%
we obtain%
\begin{equation}
\begin{tabular}{lll}
$\mathcal{L}_{B}$ & $=$ & $z\varrho _{+}+\Pi +z^{-1}\varrho _{-}+X\Pi
+z^{-1}X\varrho _{-}$ \\
&  & $+\Pi Y+z^{-1}\varrho _{-}Y+X\Pi Y+z^{-1}X\varrho _{-}Y+\frac{1}{2}%
z^{-1}X^{2}\varrho _{-}$ \\
&  & $+\frac{1}{2}z^{-1}\varrho _{-}Y^{2}+\frac{1}{2}z^{-1}X^{2}\varrho
_{-}Y $ \\
&  & $+\frac{1}{2}z^{-1}X\varrho _{-}Y^{2}+\frac{1}{4}z^{-1}X^{2}\varrho
_{-}Y^{2}$%
\end{tabular}%
\end{equation}%
To determine the matrix representation of the L-operator, we use the
following trick (projector basis)%
\begin{equation}
\mathcal{L}_{B}=\left(
\begin{array}{ccc}
\varrho _{+}L\varrho _{+} & \varrho _{+}L\Pi & \varrho _{+}L\varrho _{-} \\
\Pi L\varrho _{+} & \Pi L\Pi & \Pi L\varrho _{-} \\
\varrho _{-}L\varrho _{+} & \varrho _{-}L\Pi & \varrho _{-}L\varrho _{-}%
\end{array}%
\right)
\end{equation}%
By substituting $X=b^{i}X_{i}$ and $Y=c_{i}Y^{i}$, we end up with the
following result
\begin{equation}
\mathcal{L}_{B}^{\mu _{1}}=\left(
\begin{array}{ccc}
z+\mathbf{b}^{T}\mathbf{.c}+\frac{1}{4}z^{-1}\mathbf{b}^{2}\mathbf{c}^{2} &
\mathbf{b}^{T}-\frac{1}{2}z^{-1}\mathbf{b}^{2}\mathbf{c}^{T} & \frac{1}{2}%
z^{-1}\mathbf{b}^{2} \\
\mathbf{c}-\frac{1}{2}z^{-1}\mathbf{c}^{2}\mathbf{b} & I_{2N-1}+z^{-1}%
\mathbf{b}^{T}\mathbf{.c} & -z^{-1}\mathbf{b} \\
\frac{1}{2}z^{-1}\mathbf{c}^{2} & -z^{-1}\mathbf{c}^{T} & z^{-1}%
\end{array}%
\right)  \label{LB}
\end{equation}%
where $\mathbf{f}^{T}$ stands for the row vector ($f_{1},...,f_{2N-1}$).
Notice the two following features regarding eq(\ref{LB}). \newline
$\left( \mathbf{1}\right) $ By multiplication of $\mathcal{L}_{B}^{\mu _{1}}$
by z, we recover the Lax-matrix of B-type obtained in \cite{1E} by using
anti-dominant shifted Yangians. \newline
$\left( \mathbf{2}\right) $ Eq(\ref{LB} has a quite similar form to the Lax
operator $\mathcal{L}_{D}^{{\small vec}}$ of the $SO_{2N}$ family given by
eq(\ref{Dn1}). The main difference concerns the number of $\left( b,c\right)
$ oscillators that appears in the middle block. For $\mathcal{L}_{D_{N+1}}^{%
{\small vec}}$, we have $2N$ oscillators versus $2N-1$ oscillators for $%
\mathcal{L}_{B_{N}}.$ This property can be explained by the fact that the B$%
_{N}$ Dynkin diagram can be obtained from the folding the two spinorial-like
nodes of the D$_{N+1}$ Dynkin diagram as depicted by the Figure \textbf{\ref%
{fol}. }In this folding, the vectorial minuscule coweight is preserved.%
\textbf{\ }
\begin{figure}[tbph]
\begin{center}
\includegraphics[width=8cm]{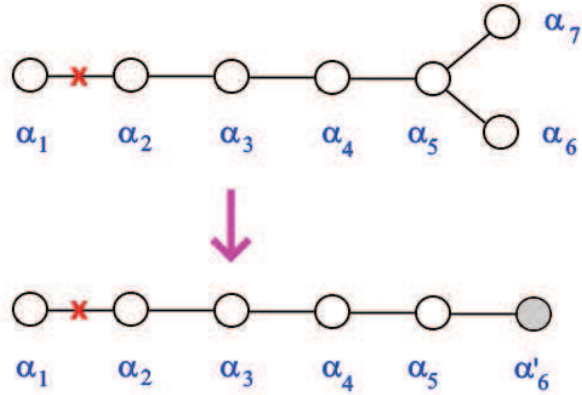}
\end{center}
\par
\vspace{-0.5cm}
\caption{Folding the nodes $\protect \alpha _{N-1}$ and $\protect \alpha _{N}$%
\ of the Dynkin Diagram of D$_{N+1}$ giving the Dynkin diagram of B$_{N}$.
This folding commutes with the cutting of the node $\protect \alpha _{1}$.}
\label{fol}
\end{figure}

\section{C-type Lax operators}

In this section, we derive the minuscule Lax operator $\mathcal{L}_{C}(z)$\
from the 4D Chern- Simons theory with gauge symmetry $SP_{2N}.$ The $%
\mathcal{L}_{C}^{\mu }$ is determined by using the formula $e^{X}z^{\mu
}e^{Y}$. Here, the $\mu $ is the minuscule coweight of the Lie algebra $%
C_{N}\sim sp_{2N}$ and the $X$ and $Y$ are $2N\times 2N$ matrices belonging
to the nilpotent sub-algebras $\boldsymbol{n}_{-}$ and $\boldsymbol{n}_{+}$
appearing in the Levi-decomposition of $sp_{2N},$ namely%
\begin{equation}
\begin{tabular}{ccc}
$sp_{2N}$ & $=$ & $\boldsymbol{l}_{\mu }\oplus \boldsymbol{n}_{+}\oplus
\boldsymbol{n}_{-}$ \\
$\boldsymbol{l}_{\mu }$ & $=$ & $so_{2}\oplus sl_{N}$ \\
$\boldsymbol{n}_{\pm }$ & $=$ & $N_{\pm 1}$%
\end{tabular}
\label{c1}
\end{equation}%
The dimensions and the ranks of the algebras involved in this decomposition
are as given below%
\begin{equation}
\begin{tabular}{ccccc}
$\dim $ & : & $N\left( 2N+1\right) $ & $=$ & $1+\left( N%
{{}^2}%
-1\right) +\frac{1}{2}N\left( N+1\right) $ \\
$rank$ & : & $N$ & $=$ & $1+\left( N-1\right) $%
\end{tabular}%
\end{equation}%
Recall that the Lie algebra $C_{N}$ has one minuscule coweight reading in
terms of the $\left \{ e_{i}\right \} $ weight vector basis as $\mu =\mu
_{N}=\frac{1}{2}\left( e_{1}+...+e_{N}\right) $. This minuscule coweight is
the dual of the simple root $\alpha _{N}=2e_{N}$, it corresponds to the N-th
node of the Dynkin diagram of $C_{N}$ given by the Figure \textbf{\ref{C6}}.%
\newline
Recall also that in (\ref{c1}), $\boldsymbol{l}_{\mu }$ is the
Levi-subalgebra of $sp_{2N}$ and the $\boldsymbol{n}_{\pm }$ are the
nilpotent sub-algebras having $\frac{1}{2}N\left( N+1\right) $ dimensions
that split like $\frac{1}{2}N\left( N-1\right) +N$. These subalgebras are
generated by matrix generators denoted like $\left( X_{\left[ i\bar{j}\right]
},X_{i\bar{\imath}}\right) $ and $\left( Y^{\left[ \bar{\imath}j\right] },Y^{%
\bar{\imath}i}\right) .$ They obey the following commutation relations
\begin{equation}
\begin{tabular}{lll}
$\left[ \mathbf{\mu },X_{i\bar{\imath}}\right] $ & $=$ & $+X_{i\bar{\imath}}$
\\
$\left[ \mathbf{\mu },X_{\left[ i\bar{j}\right] }\right] $ & $=$ & $+X_{%
\left[ i\bar{j}\right] }$ \\
$\left[ \mathbf{\mu },Y^{\bar{\imath}i}\right] $ & $=$ & $-Y^{\bar{\imath}i}$
\\
$\left[ \mathbf{\mu },Y^{\left[ \bar{\imath}j\right] }\right] $ & $=$ & $-Y^{%
\left[ \bar{\imath}j\right] }$%
\end{tabular}
\label{lde}
\end{equation}%
where $\mathbf{\mu }$\ stands for the adjoint action of the minuscule
coweight.

\subsection{Solving Levi-constraints for C$_{N}$}

To solve\ the constraint relations (\ref{lde}), we need the
Levi-decomposition of the fundamental representation $\mathbf{2N}$ of the
Lie algebra $sp_{2N}$ given by
\begin{subequations}
\begin{equation}
\mathbf{2N}=\mathbf{N}_{+1/2}\oplus \mathbf{N}_{-1/2}\equiv \mathbf{N}\oplus
\mathbf{\bar{N}}
\end{equation}%
where $\mathbf{N}_{+1/2}$ and $\mathbf{N}_{-1/2}$ are representations of sl$%
_{N}$ and the subscripts $\pm 1/2$ referring to the SO$_{2}$ charges. To
proceed, we use $\left( \mathbf{a}\right) $ the 2N kets $\left \{
\left
\vert i\right \rangle ,\left \vert \bar{\imath}\right \rangle
\right
\} $ with $i=1,...,N$ and $\bar{\imath}=2N+1-i$ to represent the
quantum basis states of the symplectic representation $\mathbf{2N}$. For
convenience, we order the $\bar{\imath}$-label like $\bar{\imath}=\bar{N}%
,...,\bar{1}$. $\left( \mathbf{b}\right) $ the dual bras $\left \{
\left
\langle i\right
\vert ,\left \langle \bar{\imath}\right \vert
\right
\} $ to denote the vector basis of ($\mathbf{2N}$)$^{T}$. Formally,
we have
\end{subequations}
\begin{equation}
\left \vert \mathbf{2N}\right \rangle =\left(
\begin{array}{c}
\left \vert i\right \rangle \\
\left \vert \bar{\imath}\right \rangle%
\end{array}%
\right) \qquad ,\qquad \left \langle \mathbf{2N}\right \vert =\left(
\begin{array}{c}
\left \langle i\right \vert \\
\left \langle \bar{\imath}\right \vert%
\end{array}%
\right)
\end{equation}%
The next step is to work out the adjoint action $\mathbf{\mu }$ of the
minuscule coweight on the fundamental representation $\mathbf{2N}$. It is
given by
\begin{equation}
\mathbf{\mu }=\frac{1}{2}\Pi _{_{\mathbf{N}}}-\frac{1}{2}\Pi _{_{\mathbf{%
\bar{N}}}}
\end{equation}%
with the projectors on $\mathbf{N}_{+1/2}$ and $\mathbf{N}_{-1/2}$ as follows%
\begin{equation*}
\Pi _{_{\mathbf{N}}}=\sum_{i=1}^{N}\varrho _{i}\qquad ,\qquad \Pi _{_{%
\mathbf{\bar{N}}}}=\sum_{\bar{\imath}=\bar{1}}^{\bar{N}}\bar{\varrho}_{i}
\end{equation*}%
In these relations, we have set
\begin{equation}
\begin{tabular}{lll}
$\varrho _{i}$ & $=$ & $\left \vert i\right \rangle \left \langle i\right
\vert $ \\
$\bar{\varrho}_{i}$ & $=$ & $\left \vert \bar{\imath}\right \rangle \left
\langle \bar{\imath}\right \vert $%
\end{tabular}%
\end{equation}%
Now, we move to the determination of explicit expressions of the matrices $%
\left( X_{\left[ i\bar{j}\right] },X_{i\bar{\imath}}\right) $ and $\left( Y^{%
\left[ \bar{\imath}j\right] },Y^{\bar{\imath}i}\right) $ generating the
nilpotent algebras $\boldsymbol{n}_{+}$ and $\boldsymbol{n}_{-}$. They are
obtained by solving the Levi-constraint relations (\ref{lde}), we have found%
\begin{equation}
\begin{tabular}{lll}
$X_{i\bar{\imath}}$ & $=$ & $\left \vert i\right \rangle \left \langle \bar{%
\imath}\right \vert $ \\
$X_{\left[ i\bar{j}\right] }$ & $=$ & $\left \vert i\right \rangle \left
\langle \bar{j}\right \vert -\left \vert j\right \rangle \left \langle \bar{%
\imath}\right \vert $%
\end{tabular}
\label{xy}
\end{equation}%
and%
\begin{equation}
\begin{tabular}{lll}
$Y^{\bar{\imath}i}$ & $=$ & $\left \vert \bar{\imath}\right \rangle \left
\langle i\right \vert $ \\
$Y^{\left[ \bar{\imath}j\right] }$ & $=$ & $\left \vert \bar{\imath}\right
\rangle \left \langle j\right \vert -\left \vert \bar{j}\right \rangle \left
\langle i\right \vert $%
\end{tabular}
\label{yx}
\end{equation}%
They also obey other useful properties such as $[X_{i\bar{\imath}},X_{\left[
k\bar{l}\right] }]=0$ and $\left[ Y^{\bar{\imath}i},Y^{\left[ \bar{k}l\right]
}\right] =0.$ With the generators (\ref{xy}-\ref{yx}), we can express the X
and Y matrices appearing in the Lax operator. We have
\begin{equation}
\begin{tabular}{lll}
$X$ & $=$ & $b^{i\bar{\imath}}X_{i\bar{\imath}}+b^{\left[ i\bar{j}\right]
}X_{\left[ i\bar{j}\right] }$ \\
$Y$ & $=$ & $c_{\bar{\imath}i}Y^{\bar{\imath}i}+c_{\left[ \bar{\imath}j%
\right] }Y^{\left[ \bar{\imath}j\right] }$%
\end{tabular}%
\end{equation}%
In these expansions, the $\left \{ b^{i\bar{\imath}},b^{\left[ i\bar{j}%
\right] }\right \} $ and the $\left \{ c_{\bar{\imath}i},c_{\left[ \bar{%
\imath}j\right] }\right \} $ variables are the phase space coordinates of
the $\mathcal{L}_{C} $-operator.

\subsection{Building the operator $\mathcal{L}_{C}$}

\begin{figure}[tbph]
\begin{center}
\includegraphics[width=8cm]{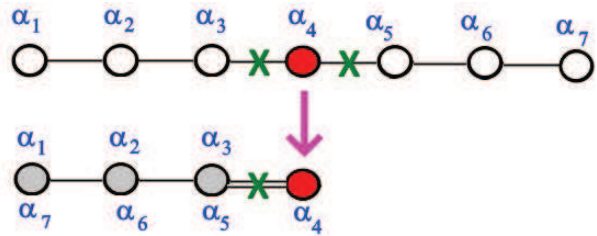}
\end{center}
\par
\vspace{-0.5cm}
\caption{Folding the nodes $\protect \alpha _{2M-1-i}$ and $\protect \alpha %
_{i}$\ of the Dynkin Diagram of A$_{2M-1}$ giving the Dynkin diagram of C$%
_{M}$. This folding commutes with the cutting of the node $\protect \alpha %
_{M}$. The green cross ($\times $) refers to the cutting of the node $%
\protect \alpha _{M}$. The folding is illustrated for the example M=4.}
\label{C4}
\end{figure}

First, we use eqs(\ref{xy}-\ref{yx}) to determine the powers $X$ and $Y.$
Then, we calculate the exponentials $e^{X}$ and $e^{Y}$ appearing in the
L-operator formula. Straightforward algebra leads to%
\begin{equation}
\begin{tabular}{lllllll}
$X_{i\bar{\imath}}X_{j\bar{j}}$ & $=$ & $0$ & \qquad ,\qquad & $Y^{\bar{%
\imath}i}Y^{\bar{j}j}$ & $=$ & $0$ \\
$X_{\left[ i\bar{j}\right] }X_{\left[ k\bar{l}\right] }$ & $=$ & $0$ &
\qquad ,\qquad & $Y^{\left[ \bar{\imath}j\right] }Y^{\left[ \bar{k}l\right]
} $ & $=$ & $0$ \\
$X_{i\bar{\imath}}X_{\left[ k\bar{l}\right] }$ & $=$ & $0$ & \qquad ,\qquad
& $Y^{\bar{\imath}i}Y^{\left[ \bar{k}l\right] }$ & $=$ & $0$%
\end{tabular}%
\end{equation}%
These properties show that the exponentials take simple forms $%
e^{X}=I_{2N}+X $ and $e^{Y}=I_{2N}+Y.$ Then, the Lax operator expands as
follows
\begin{equation}
\mathcal{L}_{C}=\left( 1+X\right) z^{\mathbf{\mu }}\left( 1+Y\right)
\end{equation}%
reading explicitly as
\begin{equation}
\mathcal{L}_{C}=z^{\mathbf{\mu }}+z^{\mathbf{\mu }}Y+Xz^{\mathbf{\mu }}+Xz^{%
\mathbf{\mu }}Y
\end{equation}%
Replacing the charge operator $z^{\mathbf{\mu }}$ with
\begin{equation}
z^{\mathbf{\mu }}=z^{\frac{1}{2}}\Pi +z^{-\frac{1}{2}}\bar{\Pi}
\end{equation}%
we get the following expression of the Lax operator for the symplectic family%
\begin{equation}
\begin{tabular}{lll}
$\mathcal{L}_{C}$ & $=$ & $z^{\frac{1}{2}}\Pi +z^{-\frac{1}{2}}\bar{\Pi}$ \\
&  & $+\left( z^{\frac{1}{2}}\Pi +z^{-\frac{1}{2}}\bar{\Pi}\right) Y+X\left(
z^{\frac{1}{2}}\Pi +z^{-\frac{1}{2}}\bar{\Pi}\right) $ \\
&  & $+X\left( z^{\frac{1}{2}}\Pi +z^{-\frac{1}{2}}\bar{\Pi}\right) Y$%
\end{tabular}%
\end{equation}%
This relation can be simplified by taking advantage of properties of the X
and Y matrices that descend from the generators realising (\ref{lde}). We
have $X\Pi =0$ and $\Pi Y=0$ as well $X\bar{\Pi}=X$ and $\bar{\Pi}Y=Y.$ By
substituting, we end up with%
\begin{equation}
\begin{tabular}{lll}
$\mathcal{L}_{C}$ & $=$ & $z^{\frac{1}{2}}\Pi +z^{-\frac{1}{2}}\bar{\Pi}$ \\
&  & $+z^{-\frac{1}{2}}\bar{\Pi}Y+z^{-\frac{1}{2}}X\bar{\Pi}$ \\
&  & $+z^{-\frac{1}{2}}XY$%
\end{tabular}%
\end{equation}%
In the projector basis $\left( \Pi ,\bar{\Pi}\right) $, we have the
following representation%
\begin{equation}
\mathcal{L}_{C}=\left(
\begin{array}{cc}
z^{\frac{1}{2}}\Pi +z^{-\frac{1}{2}}\Pi XY\Pi & z^{-\frac{1}{2}}\Pi X\bar{\Pi%
} \\
z^{-\frac{1}{2}}\bar{\Pi}Y\Pi & z^{-\frac{1}{2}}\bar{\Pi}%
\end{array}%
\right)  \label{1}
\end{equation}%
which also reads as%
\begin{equation}
\mathcal{L}_{C}=\left(
\begin{array}{cc}
z^{\frac{1}{2}}I_{N}+z^{-\frac{1}{2}}XY & z^{-\frac{1}{2}}X \\
z^{-\frac{1}{2}}Y & z^{-\frac{1}{2}}I_{N}%
\end{array}%
\right)  \label{LC}
\end{equation}%
In the vector basis $\left \{ \left \vert i\right \rangle ,\left \vert \bar{%
\imath}\right \rangle \right \} ,$ we have%
\begin{equation}
X=\left(
\begin{array}{ccc}
b^{1\bar{1}} & \cdots & b^{1\bar{N}} \\
\vdots & \ddots & \vdots \\
b^{N\bar{1}} & \cdots & b^{N\bar{N}}%
\end{array}%
\right) ,\qquad Y=\left(
\begin{array}{ccc}
c_{\bar{1}1} & \cdots & c_{\bar{1}N} \\
\vdots & \ddots & \vdots \\
c_{\bar{N}1} & \cdots & c_{\bar{N}N}%
\end{array}%
\right)
\end{equation}%
\ and%
\begin{equation}
XY=\left(
\begin{array}{ccc}
b^{1\bar{\imath}}c_{\bar{\imath}1} & \cdots & b^{1\bar{\imath}}c_{\bar{\imath%
}N} \\
\vdots & \ddots & \vdots \\
b^{N\bar{\imath}}c_{\bar{\imath}1} & \cdots & b^{N\bar{\imath}}c_{\bar{\imath%
}N}%
\end{array}%
\right)  \label{CXY}
\end{equation}%
Finally, notice the two following features regarding (\ref{LC}):\newline
$\left( \mathbf{1}\right) $ Eq(\ref{LC}) is, up to multiplication by z,
similar to the Lax-matrix of C-type obtained in \cite{1E} by using
anti-dominant shifted Yangians. \newline
$\left( \mathbf{2}\right) $ The obtained relations (\ref{LC}-\ref{CXY}) have
a quite similar structure as the Lax operator $\mathcal{L}_{A_{2N}}^{\mu
_{N}}$ of the $SL_{2N}$ family decomposed with respect to the fundamental
coweight $\mu _{N}$ of $SL_{2N}$. This similarity feature between $\mathcal{L%
}_{C_{N}}^{\mu _{N}}$ and $\mathcal{L}_{A_{2N}}^{\mu _{N}}$ can be explained
by the fact that the SP$_{2N}$ Dynkin diagram is related to the $SL_{2N}$
Dynkin diagram by folding the nodes $\alpha _{i}$ and $\alpha _{2N-1-i}$ as
shown on the Figure \textbf{\ref{C4}.}

\section{Conclusion and comments}

Four dimensional Chern-Simons gauge theory proposed in \cite{1A} has been\
shown to be a powerful QFT approach to deal with lower dimensional
integrable systems. Several results on integrable 1D quantum spin chains
such as the Lax operators of A- and D-types, obtained by using\ Bethe Ansatz
formalism and standard statistical physics as well as algebraic methods,
were nicely derived from the CS theory. The investigation given in this
paper is a contribution to the topological 4D CS gauge theory and its
applications. It essentially aims to complete some partial results obtained
in literature and also to gather the explicit expressions of minuscule Lax
operators $\mathcal{L}_{G}$ and classify them according to algebraic
properties of the gauge symmetry as given in the Tables \textbf{\ref{T1}}-%
\textbf{\ref{T2} }and the Table\textbf{\  \ref{T4} }given below. We recall
that from the view point of the 4D Chern-Simons theory, the $\mathcal{L}_{G}$%
's can be thought of as a matrix coupling an electrically charged Wilson
line W$_{\xi _{z}}^{\boldsymbol{R}}$ crossing a magnetically charged 't
Hooft line tH$_{\xi _{0}}^{\boldsymbol{\mu }}$. The study of this crossing
yields a general formula that corresponds to the oscillator realisation of
the Lax operator for an integrable XXX spin chain. This construction was
introduced in this paper along with the mathematical tools needed to build
our results. Among our contributions, we quote the three following: \newline
$\left( \mathbf{1}\right) $ We derived the non simply laced orthogonal B$%
_{N} $- and the symplectic C$_{N}$- families of Lax operators using the 4D
CS theory method. These calculations have not been addressed before in the
framework of the CS gauge theory. The Lax operators $\mathcal{L}_{B_{N}}$
and $\mathcal{L}_{C_{N}}$\ were calculated in section 3 with regards the
unified picture of all the $\mathcal{L}_{G}$'s. They were investigated with
further details in sections 4 and 5 and they were shown to agree with recent
expressions derived in the spin chain literature for the B$_{N}$ and C$_{N}$
symmetries. \newline
$\left( \mathbf{2}\right) $ We gave an interpretation of the links between
the B$_{N}$- and C$_{N}$-type Lax operators and their A$_{N}$- and D$_{N}$-
homologue in terms of discrete symmetries and foldings with respect to
discrete groups. We showed that these symmetries are nothing but the
outer-automorphisms of Dynkin diagrams of A$_{N}$ and D$_{N}$. These
foldings were visualized in the Figures \textbf{\ref{fol}} and \textbf{\ref%
{C4}} showing the relationships between the B$_{N}$/C$_{N}$-types and D$_{N}$%
/A$_{N}$ types.\newline
$\left( \mathbf{3}\right) $ We built the set of the minuscule Lax operators
labeled by a set parameters. In addition to the electric charge $\boldsymbol{%
R}$ of the Wilson line W$_{\xi _{z}}^{\boldsymbol{R}}$, these parameters are
given by the rank of the Lie algebra of the gauge symmetry G and its
minuscule coweights $\mu $.\ The content of this set is given by the table
\textbf{\ref{T2}}. This basic set contains five subsets: four infinite given
by the families A$_{N}$, B$_{N}$, C$_{N}$, D$_{N}$ and one finite given by
the exceptional E$_{6}$ and E$_{7}$ symmetries$.$\newline
We end this paper by giving brief comments concerning the L-operators of the
SO$_{2N}$ symmetry having a spinorial $\boldsymbol{R}$ representation with
dimension $\boldsymbol{R}=2^{N}$, this corresponds to having a Wilson line W$%
_{\xi _{z}}^{2^{N}}$ in the spinor representation crossing a 't Hooft line.
The construction of the associated L-operator $\left. \mathcal{L}%
_{D_{N}}^{\mu }\right \vert _{{\small R=2}^{{\small N}}}$ can be done by
following the same analysis that we performed in the sub-subsection 3.3.4 to
build $\left. \mathcal{L}_{D_{N}}^{\mu }\right \vert _{{\small R=2}N}$. In
fact, both the fundamental $\left. \mathcal{L}_{D_{N}}^{\mu }\right \vert _{%
{\small 2}^{{\small N}}}$ and $\left. \mathcal{L}_{D_{N}}^{\mu }\right \vert
_{{\small 2}N}$ are calculated by using the following formulas%
\begin{eqnarray}
\left. \mathcal{L}_{D_{N}}^{\mu }\right \vert _{{\small R=2}^{{\small N}}}
&=&\left. e^{X}z^{\mathbf{\mu }}e^{Y}\right \vert _{{\small R=2}^{{\small N}%
}}\qquad :Spin=2^{N}  \label{s} \\
\left. \mathcal{L}_{D_{N}}^{\mu }\right \vert _{{\small R=2}N} &=&\left.
e^{X}z^{\mathbf{\mu }}e^{Y}\right \vert _{{\small R=2}N}\text{ \ }\qquad
:Vect=2N  \label{v}
\end{eqnarray}%
However, though they look quite similar, the expressions of these two
operators are completely different, the first (\ref{s}) is realised by a $%
\nu \times \nu $ matrix with $\nu =2^{N}$, while the second (\ref{v}) is
given by a $2N\times 2N$ matrix. So, the matrix realisations of the triplets
$\left( \mathbf{\mu ,}X,Y\right) $ used in (\ref{s}) and in (\ref{v}) are
different. Below, we comment the matrix realisations of the triplet ($%
\mathbf{\mu }$,X,Y) involved in (\ref{s}). \newline
The Levi-charge $\mathbf{\mu }$ needed for the calculation of (\ref{s}) is
obtained by decomposing the $2^{N}$ representation as a direct sum of
representations of the Levi -subalgebra $\boldsymbol{l}_{\mu }$ as in eqs(%
\ref{56}-\ref{rde}). Because we have two types of $\boldsymbol{l}_{\mu }$'s
namely $\left( \mathbf{i}\right) $ $so_{2}\oplus D_{N-1}$ and $\left(
\mathbf{ii}\right) $ $so_{2}\oplus sl_{N}$, we distinguish two kinds of
reductions of $2^{N}$ with respect to $\boldsymbol{l}_{\mu }$ as shown in
the Table \textbf{\ref{T4}}
\begin{table}[h]
\centering \renewcommand{\arraystretch}{1.2} $%
\begin{tabular}{|c|c|c|c|c|c|c|}
\hline
$l_{\mu }$ & eq(\ref{56}) & eq(\ref{fge}) & $z^{\mu }$ & charge & $%
n_{+}/n_{-}$ & $L_{D_{N}}^{\mu }$ \\ \hline
$\left.
\begin{array}{c}
\text{ \ } \\
so_{2}\oplus D_{N-1} \\
\text{ \  \ }%
\end{array}%
\right. $ & $2_{+1/2}^{N-1}\oplus 2_{-1/2}^{N-1}$ & $\frac{1}{2}\varrho _{+}-%
\frac{1}{2}\varrho _{-}$ & $z^{\frac{1}{2}}\varrho _{+}+z^{-\frac{1}{2}%
}\varrho _{-}$ & $\pm \frac{1}{2}$ & $b^{i}X_{i}/c_{i}Y^{i}$ & eq(\ref{Dn4})
\\ \hline
$so_{2}\oplus sl_{N}$ & $\oplus _{n=0}^{N}N_{q_{n}}^{\wedge n}$ & $\dsum
\limits_{n=0}^{N}q_{n}\varrho _{n}$ & $\dsum
\limits_{n=0}^{N}z^{q_{n}}\varrho _{n}$ & $\frac{N-2n}{2}$ & $b^{i\bar{j}%
}X_{i\bar{j}}/c_{\bar{\imath}j}Y^{\bar{\imath}j}$ & (\ref{Dn44}-\ref{Dn5})
\\ \hline
\end{tabular}%
$%
\caption{Minuscule D$_{N}$-type Lax matrices describing the coupling of a 't
Hooft line tH$_{\protect \xi _{0}}^{\protect \mu }$ with magnetic charge $%
\protect \mu $ crossing a Wilson line W$_{\protect \xi _{z}}^{R}$ with
representation $R$ given by the spinorial 2$^{N}$ of the SO$_{2N}$ gauge
symmetry.}
\label{T4}
\end{table}
where the $\varrho _{i}$'s are projectors that can be thought of as $%
\left
\vert i\right \rangle \left \langle i\right \vert .$ Notice that in
the first row of the table, the projectors act like $\varrho _{\pm
}:2^{N}\rightarrow 2_{\pm 1/2}^{N-1}$ while in the second row, they act as $%
\varrho _{n}:2^{N1}\rightarrow \mathbf{N}_{q_{n}}^{\wedge n}$. Notice also
that the quantities $\mathbf{N}^{\wedge n}$ with powers $0\leq n\leq N$ are
the wedge product of n representations $\mathbf{N}$ of $sl_{N}$. The
dimension of each $\mathbf{N}^{\wedge n}$ is given by $N!/n!(N-n)!$. For
example, the $\mathbf{N}^{\wedge 2}$ is given by $\mathbf{N\wedge N}$ with
dimension $\frac{1}{2}N\left( N-1\right) $. The subscripts $q_{n}$ refer to
the charges under $so_{2}$ and their trace must vanish as in (\ref{rde});
that is, $\sum_{n=0}^{N}\frac{N!}{n!(N-n)!}q_{n}=0$ reading also like
\begin{equation}
\sum_{n=0}^{[N/2]}\frac{N!}{n!(N-n)!}\left( q_{n}+q_{N-n}\right) =0
\label{n2}
\end{equation}%
which is solved by taking $q_{N-n}=-q_{n},$ in particular $q_{N}=q_{0}$ and $%
q_{N-1}=q_{1}$. The value of $[N/2]$ depends on the parity of the integer N.
For even $N=2M$, the charge $q_{M}=0.$\newline
For the vector Levi-decomposition with Levi-subalgebra $\boldsymbol{l}_{\mu
}=so_{2}\oplus D_{N-1}$, the structure of the Lax matrix has $2^{2}=4$
blocks: two diagonal and two off-diagonal ones. It reads as follows%
\begin{equation}
\left. \mathcal{L}_{D_{N}}^{\mu _{1}}\right \vert _{{\small R=2}^{{\small N-1%
}}}=\left(
\begin{array}{cc}
z^{\frac{1}{2}}I_{2^{N-2}}+z^{-\frac{1}{2}}BC & z^{-\frac{1}{2}}B \\
z^{-\frac{1}{2}}C & z^{-\frac{1}{2}}I_{2^{N-2}}%
\end{array}%
\right)  \label{Dn4}
\end{equation}%
where $B$ and $C$ are $\eta \times \eta $ matrix oscillators with order $%
\eta =2^{N-1}$. They read in terms of the nilpotent generators $%
X_{i}=\left
\vert \beta _{+}\right \rangle \Gamma _{i}^{\beta _{+}\gamma
_{-}}\left
\langle \gamma _{-}\right \vert $ and $Y^{i}=\left \vert \gamma
_{-}\right
\rangle \Gamma _{\gamma _{-}\beta _{+}}^{i}\left \langle \beta
_{+}\right
\vert $ solving eq(\ref{lce}) like $B=b^{i}X_{i}$ and $%
C=c_{i}Y^{i} $. Here, the $\Gamma _{i}$s are Gamma matrices of $D_{N}$
satisfying the Clifford algebra in 2N dimensions. \newline
Regarding the Levi-decomposition with $\boldsymbol{l}_{\mu }=so_{2}\oplus
A_{N-1}$, the structure of the associated Lax- matrix has $\left( N+1\right)
^{2}$ blocks; $N+1$ of them are diagonal blocks; they correspond to the $N+1$
terms involved in the following expansion%
\begin{equation}
\mathbf{2}^{N}=\mathbf{1}_{q_{0}}\oplus \mathbf{N}_{q_{1}}\oplus \mathbf{N}%
_{q_{2}}^{\wedge 2}\oplus \cdots \oplus \mathbf{N}_{q_{n}}^{\wedge n}\oplus
\cdots \oplus \mathbf{N}_{q_{K}}^{\wedge N}  \label{ex}
\end{equation}%
For the example of $N=4$ corresponding to a 4D CS gauge theory with $SO_{8}$
gauge symmetry in the presence of a Wilson line $W_{\xi _{z}}^{R}$ with $%
\boldsymbol{R}=2^{4}$, the reduction of the spinor representation $2^{4}=16$
with respect of the Levi subalgebra $so_{2}\oplus sl_{4}$ reads as $%
1_{+2}+4_{+1}+6_{0}+4_{-1}+1_{-2}.$ From this reduction, we learn that the
adjoint action of $\mathbf{\mu }$ reads in terms of the 5 projectors as $%
z^{\pm 2}\varrho _{{\small 1}_{\pm }}+z^{\pm 1}\varrho _{{\small 4}_{\pm
}}+\varrho _{{\small 6}_{0}}.$ This indicates that the Lax-operator is 16$%
\times $16 matrix having 5 diagonal blocls as follows%
\begin{equation}
L=\left(
\begin{array}{ccccc}
\varrho _{{\small 1}_{+}}\mathcal{L}\varrho _{{\small 1}_{+}} & \varrho _{%
{\small 1}_{+}}\mathcal{L}\varrho _{{\small 4}_{+}} & \varrho _{{\small 1}%
_{+}}\mathcal{L}\varrho _{{\small 6}_{0}} & \varrho _{{\small 1}_{+}}%
\mathcal{L}\varrho _{{\small 4}_{-}} & \varrho _{{\small 1}_{+}}\mathcal{L}%
\varrho _{{\small 1}_{-}} \\
\varrho _{{\small 4}_{+}}\mathcal{L}\varrho _{{\small 1}_{+}} & \varrho _{%
{\small 4}_{+}}\mathcal{L}\varrho _{{\small 4}_{+}} & \varrho _{{\small 4}%
_{+}}\mathcal{L}\varrho _{{\small 6}_{0}} & \varrho _{{\small 4}_{+}}%
\mathcal{L}\varrho _{{\small 4}_{-}} & \varrho _{{\small 4}_{+}}\mathcal{L}%
\varrho _{{\small 1}_{-}} \\
\varrho _{{\small 6}_{0}}\mathcal{L}\varrho _{{\small 1}_{+}} & \varrho _{%
{\small 6}_{0}}\mathcal{L}\varrho _{{\small 4}_{+}} & \varrho _{{\small 6}%
_{0}}\mathcal{L}\varrho _{{\small 6}_{0}} & \varrho _{{\small 6}_{0}}%
\mathcal{L}\varrho _{{\small 4}_{-}} & \varrho _{{\small 6}_{0}}\mathcal{L}%
\varrho _{{\small 1}_{-}} \\
\varrho _{{\small 4}_{-}}\mathcal{L}\varrho _{{\small 1}_{+}} & \varrho _{%
{\small 4}_{-}}\mathcal{L}\varrho _{{\small 4}_{+}} & \varrho _{{\small 4}%
_{-}}\mathcal{L}\varrho _{{\small 6}_{0}} & \varrho _{{\small 4}_{-}}%
\mathcal{L}\varrho _{{\small 4}_{-}} & \varrho _{{\small 4}_{-}}\mathcal{L}%
\varrho _{{\small 1}_{-}} \\
\varrho _{{\small 1}_{-}}\mathcal{L}\varrho _{{\small 1}_{+}} & \varrho _{%
{\small 1}_{-}}\mathcal{L}\varrho _{{\small 4}_{+}} & \varrho _{{\small 1}%
_{-}}\mathcal{L}\varrho _{{\small 6}_{0}} & \varrho _{{\small 1}_{-}}%
\mathcal{L}\varrho _{{\small 4}_{-}} & \varrho _{{\small 1}_{-}}\mathcal{L}%
\varrho _{{\small 1}_{-}}%
\end{array}%
\right)  \label{Dn44}
\end{equation}%
where $\mathcal{L}$ refers to $\left. \mathcal{L}_{D_{N}}^{\mu
_{N}}\right
\vert _{{\small R=2}^{{\small N}}}.$ The Lax-matrix (\ref{Dn44}%
) reads in general as follows
\begin{equation}
L_{mn}=\varrho _{m}.\mathcal{L}.\varrho _{n}  \label{Dn5}
\end{equation}%
It has N+1 diagonal blocks $n\times n$ with dimensions $\frac{N!}{n!(N-n)!}$
given by the irreducible components in the expansion (\ref{ex}). Its
explicit expression is obtained by starting from $\left. e^{X}z^{\mathbf{\mu
}}e^{Y}\right \vert _{{\small R=2}^{{\small N}}}$ and substituting $X=b^{i%
\bar{j}}X_{i\bar{j}}$ and $Y=c_{\bar{\imath}j}Y^{\bar{\imath}j}$ as well as $%
z^{\mathbf{\mu }}=\sum z^{q_{n}}\varrho _{q_{n}}$ and charges $q_{n}=\frac{N%
}{2}-n$.

\end{document}